%% file: main.tex
\documentclass[conference]{IEEEtran}
\usepackage{multirow}
\usepackage{graphicx}
\usepackage{cite}
\usepackage{pifont}

\ifCLASSINFOpdf
\else
\fi
\usepackage[font=footnotesize]{subfig}
\usepackage{xcolor}
\usepackage{pifont}
\usepackage{booktabs}
\usepackage{multirow}
\usepackage{makecell}
\usepackage{graphicx}
\usepackage{array}
\usepackage[normalem]{ulem}
\usepackage{hyperref}
\usepackage{url}
\usepackage{amsmath}
\usepackage{amssymb}
\usepackage{enumitem}
\usepackage{nicematrix}
\newcommand{\corrauthor}{\textsuperscript{\ding{41}}}
\newcommand{\ucasmark}{\textsuperscript{\ensuremath{\diamond}}}
\usepackage[available,functional]{ndssbadges}

\newcommand{\cg}[1]{\textcolor{black}{#1}}

\usepackage{etoolbox}
\appto{\normalsize}{%
  \setlength{\abovedisplayskip}{2pt plus 1pt minus 2pt}%
  \setlength{\belowdisplayskip}{2pt plus 1pt minus 2pt}%
  \setlength{\abovedisplayshortskip}{1pt plus 1pt}%
  \setlength{\belowdisplayshortskip}{2pt plus 1pt minus 1pt}%
}

\begin{document}

\bstctlcite{BSTcontrol}

\input{envirs}

%

\title{Versatile yet Efficient Network Traffic Analysis:\makebox[0.33cm]{}\\Offloading Network Foundation Model to SmartNIC}


\author{
  \IEEEauthorblockN{
  Chungang Lin\IEEEauthorrefmark{1}\ucasmark,
  Xuying Meng\IEEEauthorrefmark{1}\IEEEauthorrefmark{2}\IEEEauthorrefmark{10}\corrauthor,
  Tianyu Zuo\ucasmark,
  Weiyao Zhang\IEEEauthorrefmark{1},
  Meng Shen\IEEEauthorrefmark{3},
  Ruijie Zhao\IEEEauthorrefmark{4}, \\
  Guanming Che\IEEEauthorrefmark{5},
  Ruiqi Meng\ucasmark,
  Ziyue Huang\IEEEauthorrefmark{1}\ucasmark,
  Haitong Luo\IEEEauthorrefmark{1}\ucasmark,
  Zhiwei Xu\IEEEauthorrefmark{6},
  Yujun Zhang\IEEEauthorrefmark{1}\ucasmark\textsuperscript{\#}\IEEEauthorrefmark{10}\corrauthor}

  \IEEEauthorblockA{
  \IEEEauthorrefmark{1}Institute of Computing Technology, Chinese Academy of Sciences}

  \IEEEauthorblockA{
  \ucasmark University of Chinese Academy of Sciences
  \IEEEauthorrefmark{2}Purple Mountain Laboratory}

  \IEEEauthorblockA{
  \IEEEauthorrefmark{3}Beijing Institute of Technology
  \IEEEauthorrefmark{4}Southeast University
  \IEEEauthorrefmark{5}Northeastern University, China
  \IEEEauthorrefmark{6}Haihe Lab of ITAI}

  \IEEEauthorblockA{
  \textsuperscript{\#}Nanjing Institute of InforSuperBah
  \IEEEauthorrefmark{10}University of Chinese Academy of Sciences, Nanjing
  \corrauthor Corresponding authors}
}


%


\IEEEoverridecommandlockouts
\makeatletter\def\@IEEEpubidpullup{6.5\baselineskip}\makeatother
\IEEEpubid{\parbox{\columnwidth}{
    Network and Distributed System Security (NDSS) Symposium 2027\\
    22 – 26 March 2027, Seoul, Republic of Korea\\
    ISBN 978-1-970672-09-1\\
    https://dx.doi.org/10.14722/ndss.2027.231231\\
    www.ndss-symposium.org
}
\hspace{\columnsep}\makebox[\columnwidth]{}}

\maketitle

\input{Abstract}

\input{Introduction}


\input{Background}


\input{Motivation}


\input{Method}


\input{Evaluation}

\input{Discussion}

\input{Conclusion}


\input{Acknowledgment}

\bibliographystyle{IEEEtranS}
\bibliography{sample-base}

\appendices
\input{Appendix}



%



\end{document}

%% file: envirs.tex
\newcommand{\allnotes}[1]{} 

\renewcommand{\allnotes}[1]{#1} 

\newcommand{\rc}[1]{{\color{blue}{#1}}}
\newcommand{\tianyu}[1]{{\allnotes{\textcolor{blue}{[Tianyu: #1]}}}}

%% file: Abstract.tex
\begin{abstract}

Pervasive encryption makes large-scale labeling infeasible for traffic analysis, while security operations demand edge analysis to avert service degradation and vulnerabilities.
\cg{These pressures have produced two research lines that advance 1) versatile analysis, via network foundation models for low label dependency, and 2) efficient analysis, via hardware offloading for low analysis latency.}
\cg{However, co-achieving versatility and efficiency via such lines has appeared incompatible, with prior work advancing one while compromising the other, yet we show that this incompatibility is a consequence of polarized design choices across the three components of traffic analysis systems, i.e., traffic processing, model architecture, and analysis execution.}

In response, we present \textit{Nepco}, a versatile yet efficient network traffic analysis system that offloads network foundation models to SmartNIC.
\cg{
Our key observation is that discriminative traffic information often exhibits localized patterns within byte sequences, motivating versatile yet efficient localized byte-sequence modeling.}
To exploit this without incurring the latency bottlenecks of complex encoding steps, we employ a hardware-friendly processing strategy that directly embeds raw byte sequences. 
To maintain versatility across diverse tasks characterized by spatially shifting features, \cg{we explore an observation-driven convolution architecture equipped with hardware-friendly scoring and gating designs.}
By exploiting translation invariance, this design dynamically locates and extracts salient semantic signatures irrespective of their byte-level offsets.
We prototype \textit{Nepco} on the NVIDIA BlueField-3 SmartNIC with multi-engine collaborative analysis execution.
The experimental results demonstrate that \textit{Nepco} achieves macro F1 competitive with the best performances achieved by 8 state-of-the-art network foundation models,
\cg{while reducing analysis latency by 328$\times$ to the millisecond scale under the SmartNIC deployment setting.}
\cg{With only 1\% labels, \textit{Nepco} outperforms existing hardware-offloadable models trained on 100\% labels under the constrained early-stage analysis setting.}

\end{abstract}

%% file: Introduction.tex
\section{Introduction}

Network traffic analysis, which aims to uncover latent information from traffic, such as malware families and service types, underpins a broad range of security applications, including website fingerprinting \cite{Swallow0CCS25,Holmes-CCS24}, intrusion detection \cite{A-NIDS-TIFS25}, and malware family classification \cite{Exposing_Malware0CCS22}.
As today's networks evolve, performing effective network traffic analysis has become increasingly challenging, driven by two compounding pressures.
First, with over 95\% of network traffic now encrypted \cite{Zscaler}, obtaining the large-scale ground-truth labels has become extremely difficult \cite{HSD}.
Second, modern security operations demand low-latency analysis at the network edge with constrained resources, before attacks cause visible service degradation and further security vulnerabilities \cite{aws_edge_ddos_whitepaper,aws_ddos_mitigation_techniques}.

\begin{figure}[t]
  \centering
  \includegraphics[width=\linewidth]{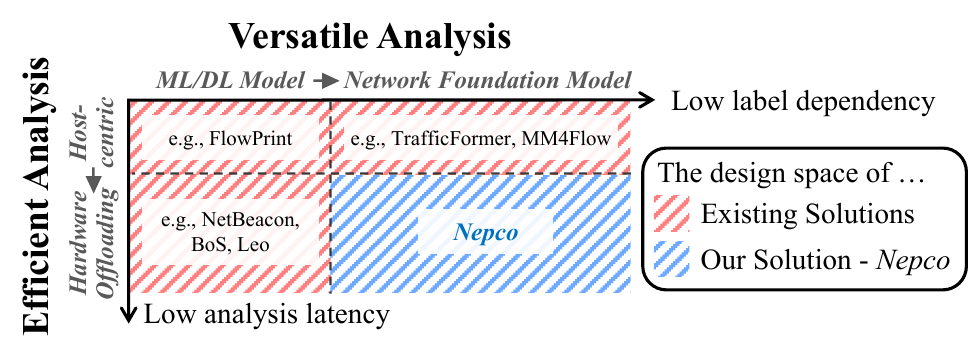}
  \caption{\cg{An illustration of the design space of existing solutions and \textit{Nepco} in network traffic analysis. Existing solutions advance versatile analysis with network foundation models (e.g., TrafficFormer, MM4Flow), or advance efficient analysis with hardware offloading (e.g., NetBeacon, BoS). By co-designing the traffic analysis system, \textit{Nepco} is presented as a hardware-offloadable network foundation model, thereby advancing versatile yet efficient network traffic analysis.}}

  \vspace{-4pt}
  \label{fig:Design-Space}
\end{figure}

These two pressures have produced two lines of research in network traffic analysis, i.e., versatile analysis and efficient analysis.
\emph{Versatile analysis} aims for models that generalize across multiple security tasks with only limited labeled data \cite{MM4Flow-CCS25}.
Inspired by the success of the pre-training and fine-tuning paradigm in developing visual and large language models \cite{BERT,GPT2,MAE,ViT}, the network and security community has increasingly explored pre-training on large-scale unlabeled network traffic to develop network foundation models (e.g., \cite{ET-BERT-WWW22,TrafficFormer-SP25,MM4Flow-CCS25,ASNet-TIFS25,Pcap-Encoder-SIGCOMM25,Flow-MAE-RAID23,Nuwa-ToN25,ATVITSC-TIFS24,NetMamba-ICNP24,YaTC-ToN24}, to name a few), supporting multi-tasking by fine-tuning with limited labeled data.
\cg{\emph{Efficient analysis}, on the other hand, aims to enable low-latency analysis under constrained resources.}
\cg{Hardware offloading advances efficient analysis by moving part or all of the analysis system onto programmable hardware, such as programmable switches \cite{P4-Switch}, SmartNICs \cite{nvidia_bluefield_suricata}, or FPGAs \cite{FPGA}, enabling on-device analysis without host involvement \cite{Poseidon-NDSS20,FlowLens-NDSS21,Jaqen-USENIX-Security21,HorusEye-USENIX-Security23,NetBeacon-USENIX-Security23,Exosphere-CCS24,MineShark-NDSS25,Whisper-CCS21,HyperVision-NDSS23,pVoxel-CCS23,tFusion-CCS25}.}

\cg{However, the versatility advanced by network foundation models and the efficiency advanced by hardware offloading have appeared incompatible. Prior work has documented sharp tradeoffs between the two.}
Naively deploying network foundation models (e.g., ET-BERT \cite{ET-BERT-WWW22}) onto SmartNICs can incur second-scale inference latency\cite{SmartTC-IWQoS25}, which is far above the sub-second budget that motivates hardware offloading \cite{aws_ddos_mitigation_techniques}.
Conversely, hardware-offloadable models cannot close the gap by scaling up training labels. 
\cg{Our measurement study in Section \ref{sec:Motivation} shows that, even when trained on 140.01$\times$ more labeled flows, two state-of-the-art hardware-offloadable models \cite{BoS-NSDI24,Flowrest-INFOCOM23} still lag behind two network foundation models \cite{ET-BERT-WWW22,TrafficFormer-SP25} by up to 54.17\% in macro F1 under the early-stage analysis setting.}
\cg{These observations appear to confirm an incompatibility between such advanced versatility and efficiency,} posing a critical obstacle for security-sensitive deployments, where defenders must detect threats and respond within tight latency budgets to prevent damage.

\definecolor{CrossRed}{HTML}{B22322}
\definecolor{CheckBlue}{HTML}{4169E1}

\newcommand{\xmark}{\textcolor{CrossRed}{\ding{55}}}
\newcommand{\cmark}{\textcolor{CheckBlue}{\ding{51}}}
\newcommand{\cmarksup}[1]{\textcolor{CheckBlue}{\ding{51}\textsuperscript{#1}}}

\newcolumntype{F}[1]{>{\centering\arraybackslash}p{#1}}

Yet a core question remains: ``\emph{Is this incompatibility fundamental, or a consequence of the design choices each line has made}?''
\cg{We show that this incompatibility is not fundamental, but a consequence of polarized design choices across the three components of network traffic analysis systems:}

\vspace{1pt}

\noindent \textbf{1) Traffic Processing.} Network foundation models adopt fine-grained semantic inputs that preserve transferable information but require computationally complex encoding, e.g., tokenization \cite{ET-BERT-WWW22,TrafficFormer-SP25}, unfriendly to edge hardware; hardware-offloadable models, conversely, adopt coarse-grained statistical inputs that are cheap to construct but lose fine-grained information, requiring prolonged traffic observation to compensate.

\vspace{1pt}

\noindent \textbf{2) Model Architecture.} Network foundation models mostly adopt expressive Transformer \cite{Transformer} architectures (e.g., BERT-style \cite{TrafficFormer-SP25,MM4Flow-CCS25}, MAE-style \cite{YaTC-ToN24,Flow-MAE-RAID23}) whose $\mathcal{O}(n^2)$ complexity with respect to sequence length precludes hardware inference; hardware-offloadable models adopt compact model architectures (e.g., Tree \cite{Leo-NSDI24}, Forest \cite{NetBeacon-USENIX-Security23}), but with limited representational capacity and high label dependency.

\vspace{1pt}

\noindent \textbf{3) Analysis Execution.} Network foundation models remain host-centric, gaining the flexibility of general-purpose compute platforms but paying substantial packet-delivery overhead before inference; hardware-offloadable models stay hardware-centric, gaining line-rate packet processing but at the cost of restricting model inference to constrained hardware primitives.

\vspace{1pt}



\cg{At each component, prior work advances one while compromising the other.}
A system co-designed across the three components, in contrast, has the potential to achieve both.

\cg{In this paper, we present \textit{Nepco}, a versatile yet efficient network traffic analysis system that advances network foundation models toward hardware offloading.}
\cg{As shown in Fig. \ref{fig:Design-Space}, prior work typically advances versatile analysis with network foundation models but remains host-centric (e.g., \cite{TrafficFormer-SP25,MM4Flow-CCS25}), or advances efficient analysis with hardware offloading but retains high label dependency (e.g., \cite{NetBeacon-USENIX-Security23,BoS-NSDI24}).}
\cg{We have also witnessed efforts to improve the efficiency of network foundation models (e.g., \cite{NetMamba-ICNP24,SmartTC-IWQoS25}), but these efforts mainly consider optimization under host-centric execution and do not necessarily translate directly into hardware-friendly designs, whereas we specifically consider hardware offloading.}


\cg{We design an encoding-free early-stage traffic processing strategy that is hardware-friendly and constructs model inputs from packet byte sequences while preserving fine-grained traffic semantics using byte embeddings.}
\cg{We build on established byte embedding techniques and implement the embedding lookup as hardware table lookups, eliminating the computationally complex encoding steps (e.g., tokenization) required by existing network foundation models.}
\cg{The preserved fine-grained semantics enable early-stage analysis from the first few packets, unlike existing hardware-offloadable models that rely on coarse-grained statistical inputs and thus require prolonged traffic observation (up to 2048 packets \cite{BoS-NSDI24,NetBeacon-USENIX-Security23}). 
}
Additionally, prior work has raised evaluation validity concerns about address information leakage \cite{Pcap-Encoder-SIGCOMM25,PTU-ICNP24}.
We therefore zero out address-related protocol fields.
As a concrete illustration, the macro F1 of Pcap-Encoder \cite{Pcap-Encoder-SIGCOMM25} changes from 0.5020 to 0.9008 depending on whether this step is applied.

\cg{We design an observation-driven convolution-based network foundation model architecture with $\mathcal{O}(n)$ complexity, equipped with dedicated hardware-friendly window-wise byte scoring and sequence-wise byte gating for localized pattern-aware traffic modeling.
While Transformers are the \textit{de facto} choice for existing network foundation models, our analysis in Section \ref{sec:Localized Pattern-Aware Traffic Modeling} shows that discriminative traffic information often exhibits localized patterns within byte sequences.
This motivates us to explore a convolution-based architecture that efficiently captures such patterns.
We build on established depth-wise convolution \cite{Xception-CVPR17} for its channel-parallel computation well-suited for hardware execution, and further tailor it to the observed localized traffic patterns through window-wise byte scoring and sequence-wise byte gating, which emphasize informative bytes within local windows and across the byte sequence, respectively, while retaining hardware-friendly execution through pre-computable scoring weights and a parallel depth-wise convolution branch with element-wise operations.}


We implement a prototype of \textit{Nepco} on the NVIDIA BlueField-3 SmartNIC, and evaluate \textit{Nepco} extensively with six network traffic analysis tasks.
\cg{We introduce a SmartNIC-native multi-engine collaborative execution strategy that coordinates the programmable NIC hardware and SoC cores to keep the entire analysis system on-device, while leveraging established packet-processing and asynchronous mechanisms.
The NIC filters later-stage packets of already-analyzed flows at line rate, reducing the SoC’s processing load.
On the SoC side, input construction and model inference run in parallel with asynchronous filter-rule updates back to the NIC.
We further prototype three existing network foundation models \cite{ET-BERT-WWW22,TraGe-IWQoS25,TrafficFormer-SP25} on the same SmartNIC.
Under the SmartNIC deployment setting, \textit{Nepco} achieves macro F1 competitive with state-of-the-art network foundation models while reducing analysis latency by up to 328$\times$ to the millisecond scale.}

The core contributions of this paper are three-fold:

\begin{itemize}

    \item \cg{We show that the incompatibility between versatile analysis with network foundation models and efficient analysis with hardware offloading is not fundamental, but a consequence of the polarized design choices in prior work.}

    \item \cg{We present \textit{Nepco}, which co-designs the three core components of the network traffic analysis system, to advance network foundation models toward hardware offloading.}
    \item We prototype \textit{Nepco} on SmartNIC, and conduct extensive evaluations to validate its versatility and efficiency.
\end{itemize}

This work does not raise any ethical issues.

%% file: Background.tex
\section{Background}
\label{sec:Background}

\subsection{Network Traffic Analysis}
\label{sec:Network Traffic Analysis}

Network traffic analysis aims to infer latent information from network traffic, such as malware families and service types, and is essential for network security and management \cite{TrafficFormer-SP25,MM4Flow-CCS25,NetLLM-SIGCOMM24}.
It supports a broad range of security applications, including website fingerprinting \cite{Swallow0CCS25,Holmes-CCS24}, malware classification \cite{Exposing_Malware0CCS22}, Internet of Things attack detection \cite{WiFinger-arXiv25}, and intrusion detection \cite{A-NIDS-TIFS25}.
Several security companies, including ENEA \cite{ENEA}, TrafficWiz \cite{TrafficWiz}, allot \cite{allot}, and LiveAction \cite{LiveAction}, are actively developing network traffic analysis technologies.
For instance, LiveAction applies machine learning to analyze encrypted network traffic, supporting threat detection and network monitoring without requiring decryption.

\begin{figure}[t]
  \centering
  \includegraphics[width=\linewidth]{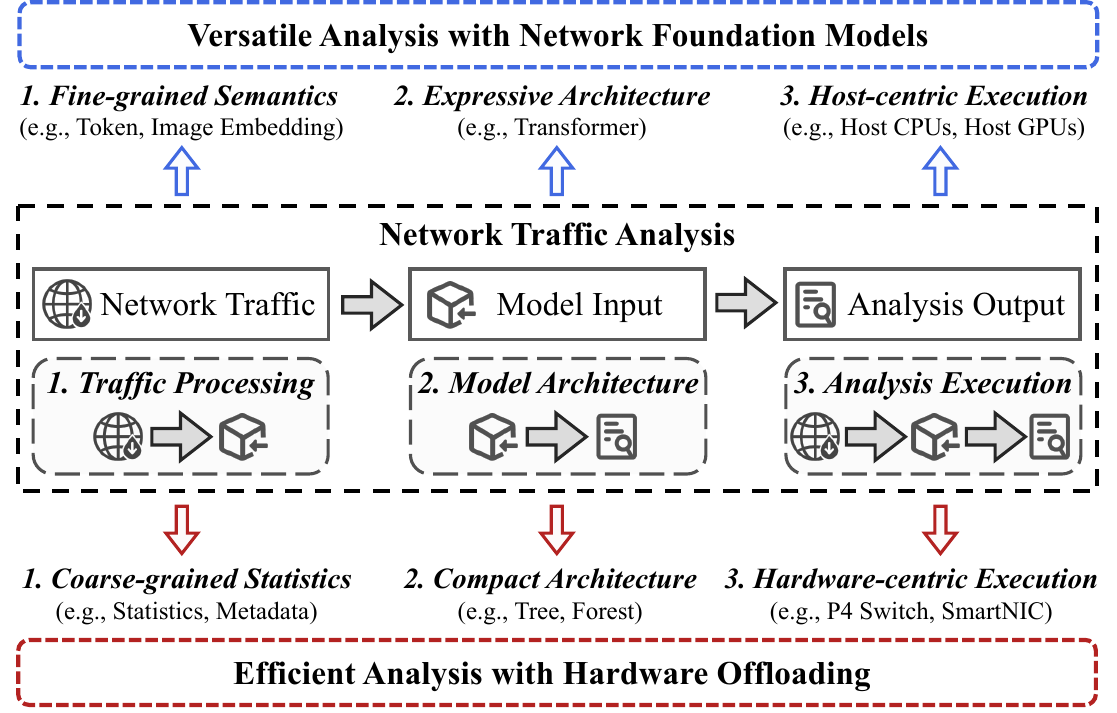}
  \caption{Conceptual decomposition of the network traffic analysis system into three core components, i.e., traffic processing, model architecture, and analysis execution, on which existing network foundation models and hardware-offloadable models make corresponding design choices to pursue versatile analysis and efficient analysis, respectively.}
  \vspace{-12pt}
  \label{fig:NTA-workflow}
\end{figure}

As shown in Fig.~\ref{fig:NTA-workflow}, the network traffic analysis system can be conceptually divided into three core components based on prior analysis pipelines \cite{CATO-NDSI25}, including \textit{\textbf{1. Traffic Processing}}, which converts captured raw network traffic into model inputs that satisfy system requirements; \textit{\textbf{2. Model Architecture}}, which is designed to capture different traffic characteristics from the model inputs, and thereby enable their transformation into final analysis outputs; and \textit{\textbf{3. Analysis Execution}}, which specifies the hardware platforms or devices on which these components are deployed and executed.

\subsection{Versatile Analysis with Network Foundation Models}
\label{sec:Versatile Analysis with Foundation Models}

Versatile analysis refers to a model’s ability to generalize across multiple downstream tasks while requiring only limited labeled data \cite{MM4Flow-CCS25,NFM_Review}. 
For instance, MM4Flow \cite{MM4Flow-CCS25} aims to obtain versatile network traffic representations to enhance multi-tasking capabilities.
This capability is increasingly important because the widespread encryption of network traffic has made large-scale labeled data difficult.
A 2024 report from Zscaler shows that 95\% of network traffic is encrypted \cite{Zscaler}, while HCSS and HSD emphasize that labeling massive volumes of network traffic data is extremely time-consuming and difficult for human annotators \cite{HSD}.

\begin{table*}[t!]
\renewcommand{\arraystretch}{0.95}
\centering
\small
\caption{Comparison of labeled data dependency between network foundation models and hardware-offloadable models.}
\label{tab:Measurement-Study}
\resizebox{1\textwidth}{!}{%
\begin{tabular}{c|c|cc|cc|cc|cc}
\toprule

\multicolumn{2}{c|}{\multirow{2}{*}{\textbf{Methods}}} &
\multicolumn{4}{c|}{\textbf{Network Foundation Models}} &
\multicolumn{4}{c}{\textbf{Hardware-Offloadable Models}} \\

\multicolumn{2}{c|}{} &
\multicolumn{2}{c|}{\textbf{ET-BERT} \cite{ET-BERT-WWW22}} &
\multicolumn{2}{c|}{\textbf{TrafficFormer} \cite{TrafficFormer-SP25}} &
\multicolumn{2}{c|}{\textbf{Flowrest} \cite{Flowrest-INFOCOM23}} &
\multicolumn{2}{c}{\textbf{BoS} \cite{BoS-NSDI24}} \\
\midrule

\multicolumn{2}{c|}{Setting of Labeled Flows \cg{\textsuperscript{1}}} &
5000 Flows & All Flows &
5000 Flows & All Flows &
5000 Flows & All Flows &
5000 Flows & All Flows \\ \midrule

\multirow{3}{*}{\makecell{\textbf{USTC-TFC (2016)} \cite{USTC-TFC-ICOIN17} \\ \textbf{All / 5000 $\approx$ 3.53}}} &
Precision &
0.9525 {\scriptsize\(\pm\ 0.0019\)} &
0.9711 {\scriptsize\(\pm\ 0.0024\)} &
0.9489 {\scriptsize\(\pm\ 0.0026\)} &
0.9731 {\scriptsize\(\pm\ 0.0033\)} &
0.7015 {\scriptsize\(\pm\ 0.0092\)} &
0.8894 {\scriptsize\(\pm\ 0.0102\)} &
0.6070 {\scriptsize\(\pm\ 0.0196\)} &
0.8166 {\scriptsize\(\pm\ 0.0205\)} \\

&
Recall &
0.9511 {\scriptsize\(\pm\ 0.0020\)} &
0.9672 {\scriptsize\(\pm\ 0.0023\)} &
0.9471 {\scriptsize\(\pm\ 0.0025\)} &
0.9698 {\scriptsize\(\pm\ 0.0038\)} &
0.7049 {\scriptsize\(\pm\ 0.0084\)} &
0.8847 {\scriptsize\(\pm\ 0.0094\)} &
0.6245 {\scriptsize\(\pm\ 0.0128\)} &
0.8007 {\scriptsize\(\pm\ 0.0217\)} \\

&
Macro F1 &
0.9509 {\scriptsize\(\pm\ 0.0020\)} &
0.9690 {\scriptsize\(\pm\ 0.0024\)} &
0.9469 {\scriptsize\(\pm\ 0.0025\)} &
0.9714 {\scriptsize\(\pm\ 0.0034\)} &
0.6869 {\scriptsize\(\pm\ 0.0078\)} &
0.8869 {\scriptsize\(\pm\ 0.0099\)} &
0.5737 {\scriptsize\(\pm\ 0.0141\)} &
0.8067 {\scriptsize\(\pm\ 0.0211\)} \\ \midrule

\multicolumn{2}{c|}{Average Performance Improvement} &
- & \textbf{$\uparrow$ 1.85\%} &
- & \textbf{$\uparrow$ 2.51\%} &
- & \textbf{$\uparrow$ 27.14\%} &
- & \textbf{$\uparrow$ 34.45\%} \\

\midrule

\multicolumn{2}{c|}{Setting of Labeled Flows} &
5000 Flows & All Flows &
5000 Flows & All Flows &
5000 Flows & All Flows &
5000 Flows & All Flows \\ \midrule

\multirow{3}{*}{\makecell{\textbf{CIC-IoMT (2024)} \cite{CIC-IoMT-arXiv24} \\ \textbf{All / 5000 $\approx$ 140.01}}} &
Precision &
0.8373 {\scriptsize\(\pm\ 0.0070\)} &
0.8323 {\scriptsize\(\pm\ 0.0081\)} &
0.8313 {\scriptsize\(\pm\ 0.0055\)} &
0.8379 {\scriptsize\(\pm\ 0.0075\)} &
0.4372 {\scriptsize\(\pm\ 0.0188\)} &
0.7268 {\scriptsize\(\pm\ 0.0198\)} &
0.1249 {\scriptsize\(\pm\ 0.0050\)} &
0.3932 {\scriptsize\(\pm\ 0.0074\)} \\

&
Recall &
0.8223 {\scriptsize\(\pm\ 0.0082\)} &
0.8202 {\scriptsize\(\pm\ 0.0090\)} &
0.8261 {\scriptsize\(\pm\ 0.0056\)} &
0.8333 {\scriptsize\(\pm\ 0.0081\)} &
0.3914 {\scriptsize\(\pm\ 0.0051\)} &
0.7188 {\scriptsize\(\pm\ 0.0192\)} &
0.2382 {\scriptsize\(\pm\ 0.0033\)} &
0.4138 {\scriptsize\(\pm\ 0.0090\)} \\

&
Macro F1 &
0.8187 {\scriptsize\(\pm\ 0.0077\)} &
0.8266 {\scriptsize\(\pm\ 0.0085\)} &
0.8235 {\scriptsize\(\pm\ 0.0053\)} &
0.8288 {\scriptsize\(\pm\ 0.0077\)} &
0.3767 {\scriptsize\(\pm\ 0.0068\)} &
0.7138 {\scriptsize\(\pm\ 0.0195\)} &
0.1434 {\scriptsize\(\pm\ 0.0024\)} &
0.3798 {\scriptsize\(\pm\ 0.0082\)} \\ \midrule

\multicolumn{2}{c|}{Average Performance Improvement} &
- & \textbf{$\uparrow$ 0.04\%} &
- & \textbf{$\uparrow$ 0.77\%} &
- & \textbf{$\uparrow$ 79.79\%} &
- & \textbf{$\uparrow$ 151.13\%} \\

\bottomrule

\multicolumn{10}{l}{\cg{\textsuperscript{1} We consider the early-stage analysis setting using the first 5 packets of each flow, which represents a constrained-setting comparison for hardware-offloadable models with longer observations.}} \\

\end{tabular}%
}
\vspace{-9pt}
\end{table*}

The ``pre-training and fine-tuning'' paradigm has been shown in the natural language processing (NLP) and computer vision (CV) communities to be effective for developing foundation models that support multiple downstream tasks while substantially reducing the dependency on labeled data \cite{BERT,GPT2,MAE,ViT,T5,Foundation_CV,ALBERT}.
Motivated by this, our community has increasingly explored adapting the ``pre-training and fine-tuning'' paradigm to network traffic analysis, \textit{leading to the development of network foundation models capable of versatile analysis} \cite{ET-BERT-WWW22,TrafficFormer-SP25,MM4Flow-CCS25,BERT-ps-TIFS25,ASNet-TIFS25,Pcap-Encoder-SIGCOMM25,Flow-MAE-RAID23,NetGPT-arXiv23,TrafficGPT-arXiv24, Nuwa-ToN25,TraGe-IWQoS25,MOTA-IWQoS25,ATVITSC-TIFS24,NetMamba-ICNP24,PASS-SECON23,YaTC-ToN24,PTU-ICNP24,Nethira-ICASSP26,Lens-arXiv24,netFound-arXiv23,IoV-BERT-IDS-TVT24,PERT-ITUK20,TraCare-TDSC26}.
Specifically, existing network foundation models first transform raw network traffic into semantically meaningful model inputs suitable for model learning, such as token \cite{ET-BERT-WWW22,TrafficFormer-SP25,NetGPT-arXiv23,MM4Flow-CCS25} or patch embeddings \cite{NetMamba-ICNP24,Flow-MAE-RAID23,YaTC-ToN24}, and then typically adopt Transformer \cite{Transformer}-based model architectures (e.g., BERT-style \cite{ET-BERT-WWW22,TrafficFormer-SP25,IoV-BERT-IDS-TVT24,TraGe-IWQoS25,ASNet-TIFS25,BERT-ps-TIFS25}, GPT2-style \cite{NetGPT-arXiv23,TrafficGPT-arXiv24}, MAE-style \cite{Flow-MAE-RAID23,YaTC-ToN24}, T5-style \cite{Pcap-Encoder-SIGCOMM25,T5-ETC}), \cg{because the self-attention mechanism has been widely proven to effectively capture global information in sequential data \cite{Transformer}, making Transformers widely adopted in existing network foundation models.}
However, due to the Transformer's quadratic computational complexity with respect to input sequence length \cite{Transformer_long_L1,Transformer_long_L2}, current network foundation models are typically deployed on host-side CPU/GPU platforms.

\noindent \textbf{Discussion:} The rapid development of network foundation models also raises concerns about evaluation validity, particularly because address leakage (e.g., MAC address and IP address) may induce shortcut learning \cite{SoK_Decoding-SP25,Pcap-Encoder-SIGCOMM25,PTU-ICNP24}.
To ensure rigorous evaluation, we adopt the following safeguards.
(1) Address information is zeroed out or removed in both pre-training and fine-tuning to prevent the model from exploiting address-related cues.
(2) The pre-training and fine-tuning datasets are kept strictly disjoint to avoid biased estimates of generalization.
(3) The fine-tuning data are kept temporally later than the pre-training data to avoid temporal bias.

\subsection{Efficient Analysis with Hardware Offloading}
\label{sec:Efficient Analysis with Hardware Offloading}

Efficient analysis, i.e., low-latency network traffic analysis under constrained resource budgets, has long been a fundamental goal in our community \cite{Poseidon-NDSS20,FlowLens-NDSS21,Jaqen-USENIX-Security21,HorusEye-USENIX-Security23,NetBeacon-USENIX-Security23,Exosphere-CCS24,MineShark-NDSS25,Whisper-CCS21,HyperVision-NDSS23,pVoxel-CCS23,tFusion-CCS25}.
This arises in operational security settings where the value of detection depends on whether the analysis-and-response loop can be completed before an attack causes visible service degradation \cite{aws_edge_ddos_whitepaper,aws_ddos_mitigation_techniques}. 
AWS reports that integrating DDoS mitigation with edge services can reduce time-to-mitigate from minutes to sub-second timescales \cite{aws_ddos_mitigation_techniques}.
The same requirement is now shaping data center security infrastructure \cite{nvidia_doca_framework,nvidia_bluefield_platform}.
In NVIDIA’s public material on the BlueField series, the platform is presented through security-facing deployments such as accelerating Suricata IDS/IPS, where packet steering and traffic inspection are offloaded from the host in order to improve analysis throughput while reducing compute overhead \cite{nvidia_bluefield_suricata}.

Hardware offloading is a mainstream approach to efficient network traffic analysis \cite{FlowLens-NDSS21,NetBeacon-USENIX-Security23,BoS-NSDI24,Mazu-SIGCOMM25,Pegasus-SIGCOMM25,FENIX-NSDI26}, as it delivers high throughput and low latency by moving part or all of the analysis pipeline onto programmable hardware, such as FPGAs \cite{FPGA}, programmable switches (P4 switches) \cite{P4-Switch}, and System-on-Chip SmartNICs (SoC SmartNICs) \cite{nvidia_bluefield_suricata}.
Prior work \cite{Poseidon-NDSS20,NetWarden-USENIX-Security20,Jaqen-USENIX-Security21,FlowLens-NDSS21} initially offloads line-rate traffic processing, e.g., feature collection and summary statistics, to such hardware, while keeping learning-based analysis on the hardware control plane \cite{Jaqen-USENIX-Security21,FlowLens-NDSS21} or the host side \cite{Poseidon-NDSS20,NetWarden-USENIX-Security20}.
More recent works further push model inference itself into FPGAs \cite{Exosphere-CCS24,N3ID-NSDI22,Traffichd-DAC24,Monemi-CC13,Tong-TPDS17} and P4 switch data planes \cite{NetBeacon-USENIX-Security23,Mousika-ToN23,Flowrest-INFOCOM23,Leo-NSDI24,BoS-NSDI24,Pegasus-SIGCOMM25}, giving rise to intelligent network dataplanes that reduce cross-plane overhead and alleviate host-side CPU/GPU load.
However, the tight constraints of programmable hardware in memory capacity, supported operators, and pipeline primitives force hardware-conscious co-design. 
As a result, existing works typically rely on readily available packet metadata (e.g., packet header fields) \cite{Mousika-ToN23,RIDS-INFOCOM24,BoS-NSDI24,Pegasus-SIGCOMM25,FENIX-NSDI26,Exosphere-CCS24} and lightweight flow statistics (e.g., packet length distribution) \cite{N3ID-NSDI22,Tong-TPDS17,Mazu-SIGCOMM25,SentinelX-INFOCOM25,Leo-NSDI24} as model inputs, and mostly adopt tree \cite{Mousika-ToN23,Leo-NSDI24,SentinelX-INFOCOM25,Tong-TPDS17}-  or forest \cite{NetBeacon-USENIX-Security23,HorusEye-USENIX-Security23,Flowrest-INFOCOM23}-based model architectures.
Shallow neural network-based approaches \cite{RIDS-INFOCOM24,Pegasus-SIGCOMM25,N3ID-NSDI22}, e.g., binary neural network (BNN) \cite{N3ID-NSDI22}, require aggressive quantization or structural rewriting, and higher-capacity deep neural networks (e.g., MAE \cite{MAE}) still often rely on FPGA assistance \cite{FENIX-NSDI26} or end-host GPU collaboration \cite{SmartTC-IWQoS25}.

\noindent \textbf{\cg{Discussion: Efficiency Beyond Hardware Offloading.}}
\cg{Many non-offloading approaches also pursue efficient network traffic analysis.}
Their techniques are aligned with the three components in Fig.~\ref{fig:NTA-workflow}, including traffic feature condensation (e.g., feature subset condensation \cite{LiM-WWW25,ABL-TC-Neurocomputing22} and key byte selection \cite{BMLP-ICC21,L-ETC-ICC23}), compact model architecture (e.g., pruning \cite{Lu-IoTJ23}, quantization \cite{Dahanayaka-CN23}, and knowledge distillation \cite{NetClus-arXiv25,NetKD-CSCWD24,MSSTRNet-CS24}), and resource-adaptive execution (e.g., early quit \cite{TaTic-ToN22,ECHO-ICNP25} and dynamic serving \cite{JITI-CoNext25,CATO-NDSI25}).
\cg{These techniques have also been applied to improve the efficiency of network foundation models \cite{NetClus-arXiv25,NetKD-CSCWD24}, laying the groundwork for optimizing individual system components.
In these non-offloading approaches, foundation model inference remains on the host side.
NVIDIA reports that host-side software processing consumes host CPU cycles \cite{nvidia_bluefield_suricata}, motivating our focus on co-designing the entire traffic analysis system for hardware offloading.}

%% file: Motivation.tex
\section{Motivation}

\label{sec:Motivation}

The primary goal is to enable versatile yet efficient network traffic analysis, driven by two practical requirements, i.e., the scarcity of large-scale labeled traffic under pervasive encryption and the  low-latency constraints of real-world security operations.
Prior work has made substantial progress along these two directions separately.
Network foundation models advance versatility, whereas hardware offloading technologies advance efficiency.
These two lines of work, however, polarize the design of the three core components of traffic analysis systems, i.e., traffic processing, model architecture, and analysis execution.
In Fig. \ref{fig:NTA-workflow}, we summarize the corresponding polarized design choices for the three core components.
As a result, these polarizations make it challenging to achieve versatile yet efficient network traffic analysis.

\textbf{Measurement Study.}
To further substantiate our analysis of these polarizations, we conduct a measurement study on label dependency.
We select two network foundation models (i.e., ET-BERT \cite{ET-BERT-WWW22} and TrafficFormer \cite{TrafficFormer-SP25}) and two hardware-offloadable models (i.e., Flowrest \cite{Flowrest-INFOCOM23} and BoS \cite{BoS-NSDI24}).
Following prior work \cite{ET-BERT-WWW22,TrafficFormer-SP25,TraGe-IWQoS25,NetGPT-arXiv23,Nethira-ICASSP26}, we consider two labeling settings, 5000 flows and all flows, indicating the maximum number of labeled flows per class.
\cg{For each flow, following existing work \cite{ET-BERT-WWW22,TrafficFormer-SP25,MM4Flow-CCS25,NetGPT-arXiv23,TraGe-IWQoS25,Nethira-ICASSP26}, we consider the first 5 packets to establish an early-stage analysis setting. This setting constrains hardware-offloadable models that are originally designed to accumulate information over longer traffic observations. We use this constrained early-stage setting to compare label dependency of different models under the same traffic observation.}
We evaluate them on a 10-class malware classification task (on USTC-TFC \cite{USTC-TFC-ICOIN17} dataset) and a 19-class Internet of Medical Things attack identification task (on CIC-IoMT \cite{CIC-IoMT-arXiv24} dataset).
To prevent the randomness caused by data sampling, we construct 20 sampled datasets for each task and each labeling setting using different random seeds, and report the results with statistical analysis. We use precision, recall, and macro F1 as evaluation metrics.

\textbf{\uline{\textit{Challenge 1: Polarization in Traffic Processing.}}}

Existing methods are pushed toward two distinct traffic processing strategies, i.e., fine-grained semantics and coarse-grained statistics. 
Network foundation models adopt fine-grained semantic inputs because such inputs preserve transferable traffic semantics beyond task-specific summary features, and can support early-stage analysis from limited observed traffic (e.g., the first 5 packets in a flow).
However, they require substantially more complex input encoding, e.g., word tokenization \cite{ET-BERT-WWW22,NetGPT-arXiv23,TrafficFormer-SP25}, n-gram construction \cite{PERT-ITUK20,TraGe-IWQoS25,TrafficFormer-SP25}, or image transformation \cite{YaTC-ToN24,NetMamba-ICNP24,Flow-MAE-RAID23}.

In contrast, existing hardware-offloadable models rely on coarse-grained statistical inputs to keep traffic processing lightweight, low-latency, and simple to construct online, but such inputs require more traffic observation to accumulate sufficient information for high-accuracy analysis \cite{NetBeacon-USENIX-Security23,Leo-NSDI24,BoS-NSDI24,Pegasus-SIGCOMM25,FENIX-NSDI26}.
For example, models such as NetBeacon \cite{NetBeacon-USENIX-Security23} and BoS \cite{BoS-NSDI24} require continuous packet observation and repeated analysis result updates as more packets arrive (i.e., up to 2048 packets \cite{NetBeacon-USENIX-Security23}), rather than making high-accuracy traffic analysis from only the first few packets.
\cg{As shown in TABLE \ref{tab:Measurement-Study}, under the constrained early-stage analysis setting on CIC-IoMT, ET-BERT and TrafficFormer achieve macro F1 scores of 0.8187 and 0.8235 with 140.01$\times$ fewer labeled flows than BoS, which only reaches 0.3798 even when trained on the all labeled-flow training set.
This contrast suggests that, under the early-stage analysis setting, additional labeled data provides limited compensation for the reduced traffic observation of hardware-offloadable models.
}

\noindent \textbf{Takeaway.}
Traffic processing is polarized, with fine-grained semantics enabling versatile analysis at the cost of complex input encoding, whereas coarse-grained statistics enable efficient analysis at the cost of prolonged traffic observation.

\textbf{\uline{\textit{Challenge 2: Polarization in Model Architectures.}}}

Model architecture fundamentally shapes the analytical capability of the analysis systems \cite{NetMamba-ICNP24,NetMamba+-arXiv26,LMLMs-NDSS26,PhishLang-arXiv24,CFGs-USENIX-Security24}.
In network traffic analysis, existing methods have gradually polarized toward two distinct architectural choices, i.e., expressive architectures and compact architectures. 
Network foundation models typically favor expressive architectures because they learn richer, more transferable traffic representations and therefore rely less on labeled data, albeit at a substantially higher computational cost.
The Transformer \cite{Transformer} exemplifies this favor because its ability to capture global dependencies has made it central to network foundation models \cite{ET-BERT-WWW22,TrafficFormer-SP25,MM4Flow-CCS25,BERT-ps-TIFS25,ASNet-TIFS25,Pcap-Encoder-SIGCOMM25,NetGPT-arXiv23,TrafficGPT-arXiv24,Nuwa-ToN25,TraGe-IWQoS25,MOTA-IWQoS25,PASS-SECON23,PTU-ICNP24,Nethira-ICASSP26,Lens-arXiv24,netFound-arXiv23,IoV-BERT-IDS-TVT24,PERT-ITUK20}, but its quadratic complexity with respect to input sequence length leads to inefficient model inference \cite{Transformer_long_L1,Transformer_long_L2}.

By contrast, hardware-offloadable models adopt compact architectures to achieve low inference overhead and efficient deployment under resource constraints.
This efficiency comes at the cost of weaker representation capacity and greater reliance on labeled data.
For instance, NetBeacon \cite{NetBeacon-USENIX-Security23} requires $\sim$ 2.6 million labeled training flows in P2P application fingerprinting tasks.
Tree \cite{Mousika-ToN23,Leo-NSDI24,SentinelX-INFOCOM25,Tong-TPDS17}-  and forest \cite{NetBeacon-USENIX-Security23,HorusEye-USENIX-Security23,Flowrest-INFOCOM23}-based architectures remain dominant in hardware-offloadable models.
Although recent studies \cite{RIDS-INFOCOM24,Pegasus-SIGCOMM25,N3ID-NSDI22,BoS-NSDI24} have explored shallow neural networks (e.g., BNN \cite{N3ID-NSDI22}, RNN \cite{BoS-NSDI24}) for hardware offloading, such models still require compression or structural transformation, such as conversion into table-matching logic on P4 switches \cite{BoS-NSDI24}, which often degrades analysis accuracy.
\cg{As shown in TABLE \ref{tab:Measurement-Study}, under the constrained early-stage analysis setting, two hardware-offloadable models, i.e., Flowrest (forest-based) and BoS (RNN-based), exhibit a strong dependence on labeled data. }
Specifically, increasing the number of labeled flows by 3.53$\times$ to 140.01$\times$ leads to average F1-score improvements of 53.47\% for Flowrest and 92.80\% for BoS.

\noindent \textbf{Takeaway.}
Model architecture is polarized, with expressive architectures improving versatility but being computationally expensive, whereas compact architectures improve efficiency but depend more heavily on labeled data.

\textbf{\uline{\textit{Challenge 3: Polarization in Analysis Execution.}}}

Analysis execution determines where packet processing and model inference are carried out in an analysis system.
The pursuit of versatility and efficiency has led to two diverging execution choices, namely host-centric execution and hardware-centric execution. 
Network foundation models usually remain host-centric \cite{YaTC-ToN24,ET-BERT-WWW22,TrafficFormer-SP25,NetMamba-ICNP24,MM4Flow-CCS25}, because expressive model architectures (e.g., Transformer) benefit from flexible software environments and general-purpose compute resources.
This execution choice, however, forces network packets to be exported from hardware to the host before inference, incurring substantial packet-delivery overhead even before analysis begins.
SmartTC \cite{SmartTC-IWQoS25} shows that merely transferring 1 MB of traffic from the SmartNIC to the host incurs millisecond-level latency, while Google Cloud estimates that full packet mirroring at 100 Gbps can generate up to \$8,046 in daily network cost \cite{googlecloud_network_pricing_packet_mirroring}.
This cost is incurred before inference, showing that host-centric execution preserves model inference flexibility at the expense of packet processing efficiency.

\begin{figure*}[t]
  \centering
  \includegraphics[width=\linewidth]{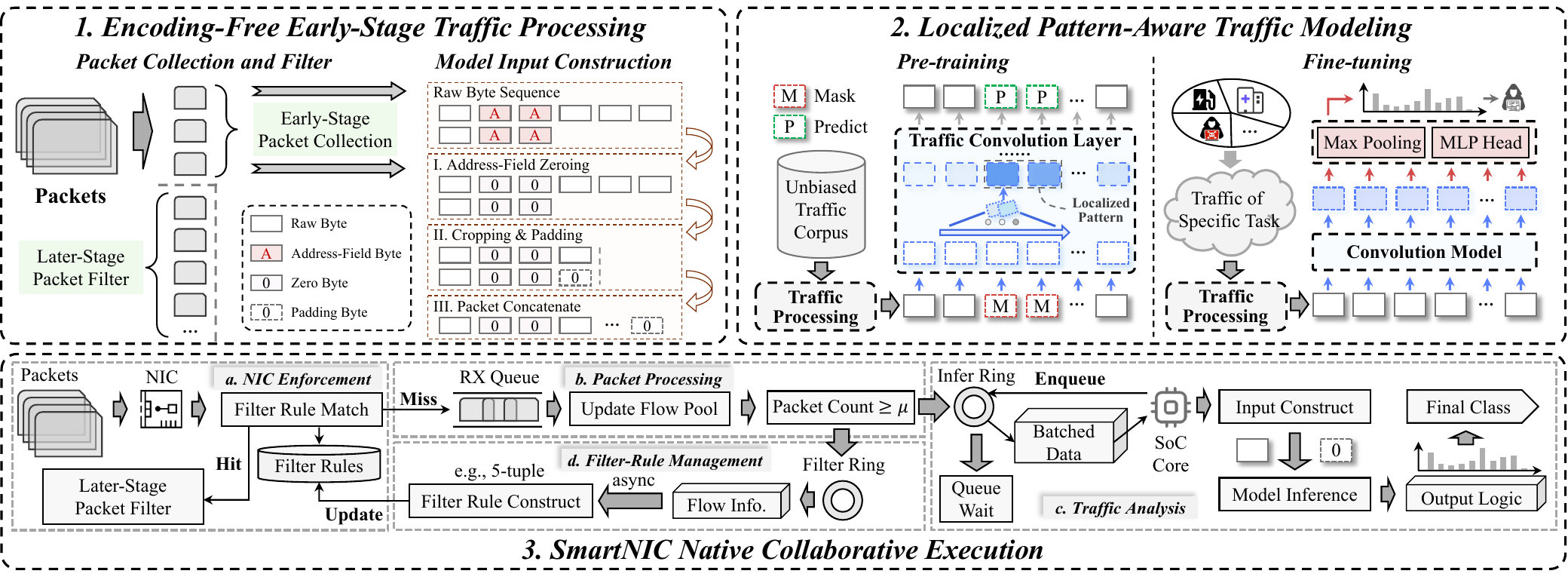}
  \caption{\cg{The overview of \textit{Nepco}.}}
  \vspace{-10pt}
  \label{fig:Overview}
\end{figure*}

Hardware-offloadable models favor hardware-centric execution because it preserves low-latency packet processing.
The cost is that model inference must conform to restricted hardware execution primitives, which directly limits its model inference flexibility.
As a result, the original inference pipeline often cannot be preserved and must be further simplified or rewritten for hardware execution.
\cg{As shown in TABLE \ref{tab:Measurement-Study}, under the constrained early-stage analysis setting, BoS, which implements its RNN inference on a P4 switch through table-matching logic, achieves a macro F1 of 0.3798 on the 19-class CIC-IoMT task, substantially lower than the 0.9230 reported for its original 3-class CIC-IoT task \cite{CIC-IoT}.
This result shows that hardware-centric execution preserves packet-processing efficiency, but may constrain the flexibility of model inference.}


\noindent \textbf{Takeaway.}
Analysis execution is polarized, with host-centric execution preserving inference flexibility at the cost of packet processing efficiency, whereas hardware-centric execution preserves processing efficiency at the cost of inference flexibility.

%% file: Method.tex
\section{Design of Nepco}
\label{sec:Design of Nepco}
\subsection{High-level Design}

We present \textit{Nepco}, a versatile yet efficient network traffic analysis system natively deployed on the SmartNIC.
\cg{As illustrated in Fig. \ref{fig:Overview}, \textit{Nepco} resolves the polarized design choices in the three components of the analysis system, i.e., traffic processing, model architecture, and analysis execution, through the three corresponding designs for hardware offloading.}



\noindent\textbf{\cg{Encoding-Free Early-Stage Traffic Processing}} \cg{resolves the polarization in traffic processing with a hardware-friendly traffic processing strategy that constructs model inputs from packet byte sequences, eliminating the complex encoding steps.
We retain the fine-grained traffic semantics necessary for versatile analysis while reducing model input construction to lightweight memory operations, and further incorporate a later-stage packet filter that drops subsequent packets of already-analyzed flows, reducing packet processing overhead.
This supports both early-stage analysis from the first few packets and unbiased model inputs through address field zeroing.}


\noindent\textbf{\cg{Localized Pattern-Aware Traffic Modeling}} \cg{resolves the polarization in model architecture by exploring the network foundation model around convolution-based traffic modeling with $O(n)$ computational complexity, compared with $O(n^2)$ for Transformer-based foundation models.
This convolution-based design is motivated by our observation that discriminative traffic information often exhibits localized patterns within byte sequences.
We follow the pre-training and fine-tuning paradigm, and tailor hardware-friendly depth-wise convolution to the observed localized traffic patterns to form traffic convolution layers for efficient traffic modeling.}


\noindent\textbf{\cg{SmartNIC Native Collaborative Execution}} \cg{resolves the polarization in analysis execution through multi-engine collaborative execution across the NIC hardware and the SoC cores of the SmartNIC.
The NIC enforcement engine performs line-rate packet filtering, forwarding unanalyzed packets to the packet processing engine, which maintains a per-flow packet pool and enqueues a flow for model inference.
The traffic analysis engine on the SoC cores then executes the input construction and the model inference entirely on-device. 
Critically, the filter rule management engine asynchronously installs the corresponding flow filter rules on the NIC, enabling later-stage packets of analyzed flows to be filtered out without re-invoking the model.
This collaborative design eliminates the packet delivery overhead of host-centric execution while preserving the model inference flexibility.}

\subsection{Overview}

\cg{
As shown in Fig. \ref{fig:Overview}, \textit{Nepco} consists of three components, corresponding to the three core components of a network traffic analysis system, i.e., traffic processing, model architecture, and analysis execution, as identified in Section \ref{sec:Network Traffic Analysis}.
}

\begin{figure}[t]
  \centering
  \includegraphics[width=1\linewidth]{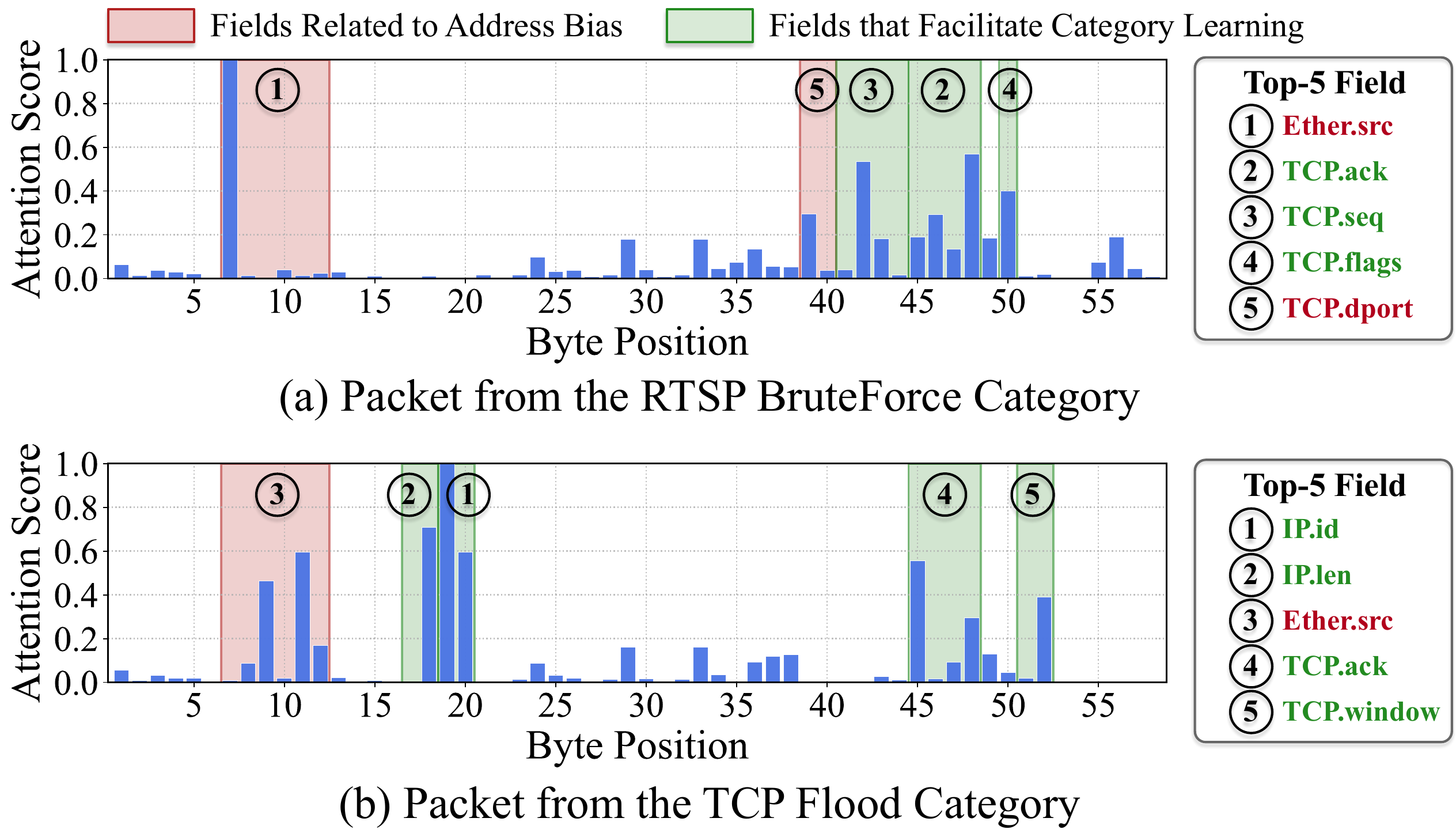}
  \caption{Byte importance pattern analysis, showing the Top-5 important protocol fields. As visualizing all traffic samples is infeasible, two representative attack traffic categories (RTSP BruteForce and TCP Flood) on the CIC-IoT dataset are used for illustration, \cg{with cross-dataset and cross-category quantitative analysis provided in Appendix \ref{app:Supporting Materials}.}}
  \vspace{-6pt}
  \label{fig:Packet-Attention}
\end{figure}

\cg{
In detail, the encoding-free early-stage traffic processing component (Section \ref{sec:Encoding-Free Traffic Data Processing}) converts network traffic into model inputs by first performing packet collection and filtering, and then constructing unbiased model inputs from the collected packets.
The localized pattern-aware traffic modeling component (Section \ref{sec:Localized Pattern-Aware Traffic Modeling}) transforms the model inputs into analysis outputs following the pre-training and fine-tuning paradigm with dedicated traffic convolution layers, whose design is motivated by the key observations in Fig. \ref{fig:Packet-Attention} and illustrated in Fig. \ref{fig:LoPA-TM}.
The SmartNIC-native collaborative execution component (Section \ref{sec:Collaborative Execution on SoC SmartNIC}) deploys and coordinates the above two components on the SmartNIC through the NIC enforcement, packet processing, filter-rule management, and traffic analysis engines, forming the end-to-end system on the SmartNIC.}

\subsection{Encoding-Free Early-Stage Traffic Processing}
\label{sec:Encoding-Free Traffic Data Processing}

\cg{We construct model inputs from packet byte sequences.
In Fig.~\ref{fig:Overview}, we first perform packet collection and filtering, and then construct unbiased model inputs from the collected packets.}

\noindent\textbf{Packet Collection and Filter.}
We collect the first $\mu$ packets of each flow for analysis.
This is motivated by the observation that the first few packets of a flow already carry sufficient semantic information for accurate analysis, as consistently demonstrated by prior work~\cite{Anderson-CCS16,Anderson-JCVHC18,Wang-ISI17,ExpMD-IWQoS24}.
We set $\mu = 5$ to align with this established setting~\cite{ET-BERT-WWW22, TrafficFormer-SP25,NetGPT-arXiv23,MM4Flow-CCS25}.
\cg{To reduce packet processing overhead, we further incorporate the Later-Stage Packet Filter.
Specifically, before processing each flow, we check whether a corresponding filter rule already exists.
Flows with an existing filter rule, i.e., flows that have been analyzed by a prior inference, are filtered without further processing.
We further detail its implementation on the NIC hardware in Section \ref{sec:Collaborative Execution on SoC SmartNIC}.}
\cg{
This filtering step is effective because it decouples the packet processing throughput from the total traffic volume; the processing overhead scales with the number of unanalyzed flows rather than the line rate, focusing traffic processing on flows that require analysis.}

\noindent\textbf{Unbiased Model Input Construction.}
\cg{For each collected flow, we construct an unbiased flat byte sequence as the model input through three sequential operations, i.e., address zeroing, cropping and padding, and packet concatenation.}
First, address fields are zeroed out at the protocol field level, covering MAC addresses at the Ethernet layer, source/destination IP addresses at the network layer, and source/destination port numbers at the transport layer.
This address zeroing is necessary because address information can induce shortcut learning \cite{Pcap-Encoder-SIGCOMM25,PTU-ICNP24,NetMamba-ICNP24,Shortcut-Nature-MI20,Comi-SIGIR24}, where the model exploits address-related cues rather than learning transferable traffic semantics, leading to biased generalization estimates.\footnote{Appendix~\ref{app:Supporting Materials} provides empirical evidence for this. Retaining address fields inflates macro F1 by 5.19\% on average, but once test-time address fields are randomized, macro F1 drops by 44.67\% on average, confirming that the gains stem from shortcut learning rather than transferable traffic semantics.}
\cg{Second, each packet's byte sequence is cropped to a fixed length of the first $128$ bytes and zero-padded if shorter, normalizing the per-packet representation to a uniform size.}
\cg{Third, the processed byte sequences of all $\mu$ packets are concatenated in arrival order to form a single flat input of $\mu \times 128$ bytes, which is directly fed into the traffic modeling component.
This preserves the inter-packet byte ordering for subsequent traffic modeling.}

\subsection{Localized Pattern-Aware Traffic Modeling}
\label{sec:Localized Pattern-Aware Traffic Modeling}


\cg{We explore a convolution-based network foundation model.
This is based on our key observation that discriminative traffic information often exhibits localized patterns within byte sequences, which motivates the design of traffic convolution layers for efficient localized pattern-aware traffic modeling.}

\subsubsection{Key Observation}

\cg{To characterize the distribution of discriminative information within network traffic,} we apply GateKeeper~\cite{GateKeeper-CN24}, a byte-level attribution method that scores each byte's contribution to the analysis decision.
Fig.~\ref{fig:Packet-Attention} illustrates the Top-5 important fields and their corresponding byte positions for two traffic categories (i.e., RTSP BruteForce and TCP Flood) on the CIC-IoT \cite{CIC-IoT} dataset.
\cg{We have the following observations that inform our design, and Appendix~\ref{app:Supporting Materials} further provides quantitative support for these observations on both in-evaluation and out-of-evaluation datasets.}


\noindent\textbf{\cg{O1: Discriminative traffic patterns are often localized.}}
\cg{As shown in Fig.~\ref{fig:Packet-Attention}, high-importance bytes tend to concentrate in a small number of protocol field regions rather than spreading uniformly across the full byte sequence, suggesting that discriminative information often exhibits localized patterns.}
Appendix~\ref{app:Supporting Materials} further supports this with right-skewed average Gini coefficients \cite{Gini} ranging from 0.2961 to 0.4939 across both in-evaluation and out-of-evaluation datasets.


\noindent\textbf{\cg{O2: Different categories favor different local regions.}}
Comparing Fig.~\ref{fig:Packet-Attention}(a) and Fig.~\ref{fig:Packet-Attention}(b), the two categories concentrate their importance on distinct protocol fields.
The RTSP BruteForce category attends primarily to transport-layer fields such as \texttt{TCP.ack} and \texttt{TCP.seq}, while the TCP Flood category attends to network-layer fields such as \texttt{IP.id} and \texttt{IP.len}.
\cg{This spatial variation across categories suggests that discriminative traffic patterns may occur at different byte-level offsets, motivating a translation-invariant model architecture to capture these patterns across varying positions.}
Appendix~\ref{app:Supporting Materials} further supports this with pairwise Jensen-Shannon Divergence \cite{JSD} between per-class field distributions consistently above zero and concentrated at moderate-to-high values.


\noindent\textbf{\cg{O3: Discriminative bytes often occupy only part of a field.}}
Within each high-importance protocol field, only a subset of bytes receives notably high scores, \cg{while the remaining bytes contribute less to the analysis.}
\cg{This variation suggests that standard convolution may not explicitly distinguish the varying importance of individual bytes, motivating selective emphasis based on byte importance.}
Appendix~\ref{app:Supporting Materials} further supports this with right-skewed average field-level Gini coefficients \cite{Gini} ranging from 0.2907 to 0.4389 across diverse datasets.

\noindent\textbf{O4: Address fields attract disproportionate importance.}
The Top-5 important fields consistently include address-related fields such as \texttt{Ether.src} and \texttt{TCP.dport}, which carry no transferable traffic semantics but receive disproportionately high importance scores.
This confirms that address zeroing is necessary not only for evaluation validity but also for directing model learning toward discriminative content.
Appendix~\ref{app:Supporting Materials} further supports this, showing that at least one address-related field appears in the Top-5 most important bytes for 88.16\% to 96.75\% of traffic samples across diverse datasets.

\subsubsection{Traffic Convolution Layer}

\cg{We adopt depth-wise convolution~\cite{Xception-CVPR17} as the base operator for its translation invariance and channel-parallel computation well-suited for hardware execution.
Existing convolution-based methods either apply convolutions to statistical features~\cite{Seq2img-Big-Data17, FlowPic-TNSM21, FlowPic-INFOCOM19, Okonkwo-ACSW22}, losing byte-level semantics, or directly apply standard convolutions to byte sequences~\cite{Wang-ISI17, Deep-Packet-SC20, CETAnalytics-CN20, ICLSTM-MDPI21, HexCNN-JCC23, Yu-CNSCT25}, without explicitly modeling the varying importance of bytes.}

\begin{figure}[t]
  \centering
  \includegraphics[width=0.9\linewidth]{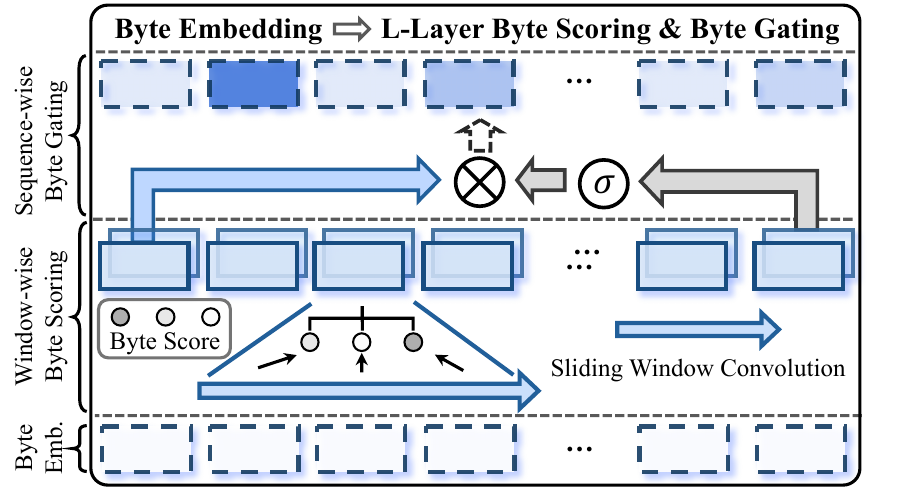}
  \caption{Architecture of the traffic modeling component.}
  \vspace{-8pt}
  \label{fig:LoPA-TM}
\end{figure}

\cg{Therefore, we tailor depth-wise convolution for localized pattern-aware traffic modeling. As shown in Fig. \ref{fig:LoPA-TM}, the input byte sequence is first mapped into a high-dimensional vector space through byte embedding, and the resulting embeddings are then processed through $L$ stacked traffic convolution layers. Each layer incorporates window-wise byte scoring and sequence-wise byte gating. Window-wise byte scoring is motivated by \textbf{O3} to emphasize informative bytes within each local convolution window, while sequence-wise byte gating is motivated by \textbf{O1} and \textbf{O2} to adaptively emphasize informative positions across the byte sequence.}

\noindent\textbf{Byte Embedding.}
The flat byte sequence consists of $\mu \times 128$ bytes.
Every two consecutive bytes are grouped into one token, yielding an input token sequence $X = (t_1, \ldots, t_N)$ of length $N = \mu \times 64$.
Each token $t_i$ is mapped to a high-dimensional embedding vector $\mathbf{e}_i$ through a learned byte embedding layer, producing the embedding sequence $(\mathbf{e}_1, \ldots, \mathbf{e}_N)$.

\cg{Unlike Transformer-based network foundation models, we do not employ explicit positional encoding, as convolution directly captures the local ordering of tokens through its sliding-window operation.}
For hardware deployment, the embedding can be implemented as a table lookup that maps each two-byte token to a pre-stored embedding vector, avoiding the large matrix multiplications required by other input constructions.

\noindent\textbf{Window-wise Byte Scoring.}
\cg{Motivated by \textbf{O3}, window-wise byte scoring assigns different importance score weights to byte positions within each local convolution window, allowing the convolution to selectively emphasize more informative positions. For each traffic convolution layer, the score weights are learned during training and shared across all sliding windows. After training, the resulting score weights are fixed and can be pre-computed and stored as constants on hardware, leaving only weighted accumulation during inference.}

For each sliding window of fixed length $k$ at layer $l$, a set of layer-specific learnable weights $\mathbf{w}^{(l)} \in \mathbb{R}^k$ is optimized to produce a score vector $\boldsymbol{\alpha}^{(l)}$, with $\sqrt{d}$ scaling to stabilize training across embedding dimensions:
\begin{equation}
    \alpha_i^{(l)} = \frac{\exp(w_i^{(l)} / \sqrt{d})}{\sum_{j=1}^{k} \exp(w_j^{(l)} / \sqrt{d})}, \quad i = 1, \ldots, k
\end{equation}
where $\alpha_i^{(l)}$ represents the importance score assigned to the $i$-th token within the window at layer $l$.
The output embedding $\mathbf{h}_t^{(l)}$ at position $t$ is computed via depth-wise convolution:
\begin{equation}
    \mathbf{h}_t^{(l)} = \sum_{i=1}^{k} \alpha_i^{(l)} \cdot \mathbf{e}_{t+i-1}^{(l)}, \quad t = 1, \ldots, N
\end{equation}
where $\mathbf{e}_{t+i-1}^{(l)} \in \mathbb{R}^d$ denotes the input embedding at position $t{+}i{-}1$ at layer $l$.
$(\mathbf{h}_1^{(l)}, \ldots, \mathbf{h}_N^{(l)})$ is an intermediate embedding sequence that captures window-level localized patterns.

\noindent\textbf{Sequence-wise Byte Gating.}
While window-wise byte scoring captures localized patterns within individual windows, certain discriminative patterns span multiple positions in the sequence and require sequence-level selectivity.
Motivated by \textbf{O1} and \textbf{O2}, sequence-wise byte gating \cg{employs} an adaptive gating mechanism~\cite{GatedCNN-ICML17} across the full sequence to emphasize informative byte positions and suppress uninformative ones.

\cg{Formally, a parallel depth-wise convolution branch with independent learnable score weights $\boldsymbol{\beta}^{(l)} \in \mathbb{R}^k$ processes the same input embeddings $\mathbf{e}^{(l)}$. At each token position $i$, the convolution output is passed through a sigmoid activation and element-wise multiplied with $\mathbf{h}_i^{(l)}$ to produce the output $\mathbf{o}_i^{(l)}$ of the $l$-th traffic convolution layer:}
\begin{equation}
    \mathbf{o}_i^{(l)} = \mathbf{h}_i^{(l)} \odot \sigma\!\left(\sum_{j=1}^{k} \beta_j^{(l)} \cdot \mathbf{e}_{i+j-1}^{(l)}\right), \quad i = 1, \ldots, N
\end{equation}
\cg{where $\sigma(\cdot)$ and $\odot$ denote the sigmoid activation and element-wise multiplication, respectively.
The output sequence $(\mathbf{o}_1^{(l)}, \ldots, \mathbf{o}_N^{(l)})$ serves as the input $\mathbf{e}^{(l+1)}$ to the next traffic convolution layer.
After stacking $L$ traffic convolution layers, $(\mathbf{o}_1^{(L)}, \ldots, \mathbf{o}_N^{(L)})$ forms the final output.}

\subsubsection{Traffic Modeling}
We adopt the pre-training and fine-tuning paradigm to learn generalizable traffic representations.

\noindent\textbf{Pre-training.}
We pre-train \textit{Nepco} using a continuous byte masking-based pre-training task, in which contiguous spans of tokens are masked~\cite{SpanBERT-TACL20}, and the model is trained to predict the original byte values from the surrounding context.
Unlike standard random byte masking~\cite{ET-BERT-WWW22, TrafficFormer-SP25, MM4Flow-CCS25}, which independently masks individual tokens, continuous byte masking explicitly trains the model to capture the structured byte patterns that correspond to protocol fields, which consist of contiguous bytes with fixed lengths and structures.

\cg{Formally, let $\mathcal{M}_{\text{mask}}$ denote the set of masked token positions. We construct $\mathcal{M}_{\text{mask}}$ by sampling contiguous token spans, where each span length $l$ is drawn from a geometric distribution. The pre-training loss is:}
\begin{equation}
    \mathcal{L} = -\frac{1}{|\mathcal{M}_{\text{mask}}|} \sum_{m \in \mathcal{M}_{\text{mask}}} \log P_\theta \!\left( t_m \mid t_1^{\text{mask}}, \ldots, t_N^{\text{mask}} \right)
\end{equation}
\cg{where $\theta$ denotes the model parameters, $t_m$ is the original token at masked position $m$, and $t_i^{\text{mask}}$ is the token at position $i$.}

\noindent\textbf{Fine-tuning.}
\cg{As shown in Fig. \ref{fig:Overview}, for each traffic analysis task, we conduct model fine-tuning with a task-specific classifier on a small set of labeled flows.
Specifically, the output sequence of the final traffic convolution layer is first aggregated into a flow-level representation through max pooling.
Max pooling selects the peak activation across token positions for each feature dimension, emphasizing the most discriminative local pattern signals in the byte sequence.
The resulting flow-level representation is then passed through a task-specific linear classifier (MLP) to produce the final analysis output.}

\subsection{SmartNIC Native Collaborative Execution}
\label{sec:Collaborative Execution on SoC SmartNIC}


\cg{We offload \textit{Nepco} to the SmartNIC to enable versatile yet efficient network traffic analysis.
As shown in Fig. \ref{fig:Overview}, four engines collaborate across the NIC hardware and the SoC cores to form the end-to-end analysis system.
The NIC enforcement engine forwards packets unmatched by the installed filter rules for early-stage collection and analysis, while filtering matched later-stage packets on the NIC hardware.
For the collected packets, the packet processing engine accumulates per-flow packet bytes and, once a flow reaches $\mu$ packets, simultaneously triggers the traffic analysis engine and the filter-rule management engine.
The traffic analysis engine executes the input construction and the model inference on-device, while the filter-rule management engine asynchronously installs the corresponding 5-tuple filter rule on the NIC.
}

\noindent\textbf{NIC Enforcement Engine.}
The NIC enforcement engine installs and enforces 5-tuple-based filter rules on the NIC hardware,
\cg{where each 5-tuple consists of the source/destination IP addresses, source/destination ports, and transport protocol.}
Each arriving packet is matched against the filter rules; packets belonging to flows already analyzed are filtered directly at the NIC without entering the SoC processing path, while unmatched packets are forwarded to the RX queue.
\cg{This design is effective because \textit{Nepco} performs early-stage analysis using the first \(\mu\) packets of each flow, following existing works \cite{ET-BERT-WWW22,TrafficFormer-SP25}. Therefore, filtering subsequent packets does not alter the input used by the deployed \textit{Nepco} configuration, while eliminating unnecessary SoC-side processing overhead.}

\noindent\textbf{Packet Processing Engine.}
The packet processing engine is implemented on the SoC cores using DPDK~\cite{DPDK}, which uses polling-mode packet reception to avoid interrupt overhead and sustain high-throughput packet processing \cite{SmartTC-IWQoS25,BoS-NSDI24}.
It maintains a per-flow packet pool indexed by a lock-free hash table keyed on the 5-tuple.
\cg{Each flow entry maintains a packet byte buffer and a packet counter.}
For each packet received from the RX queue, it performs a flow lookup, \cg{copies up to 128 bytes of the packet into the flow's byte buffer, and increments the per-flow packet counter.}
\cg{Once a flow reaches $\mu$ packets, the engine enqueues the flow entry to both the Infer Ring and the Filter Ring, triggering the traffic analysis engine and the filter-rule management engine concurrently. This parallel launch minimizes the number of later-stage packets that enter the SoC processing path before the filter rule takes effect.}

\noindent\textbf{Traffic Analysis Engine.}
The traffic analysis engine executes the complete inference pipeline on the SmartNIC without host involvement.
\cg{Flow entries wait in the Infer Ring while the current batch is under inference. Once a SoC core completes its current inference, it dequeues the next flow entry to form a new inference batch, with the batch size set to 1.
The engine then invokes the traffic processing to construct the model input from the accumulated byte buffer and runs model inference via ONNX Runtime~\cite{ONNXRuntime}.}
The resulting analysis result is written back to the flow entry as the final output. 
Unlike prior hardware offloading work that keeps model inference on the host side~\cite{SmartTC-IWQoS25}, the traffic analysis engine runs model inference entirely on-device, eliminating the packet delivery overhead that is otherwise incurred before inference even begins.

\noindent\textbf{Filter-Rule Management Engine.}
The filter-rule management engine runs on a SoC core, operating asynchronously from the traffic analysis engine.
\cg{It continuously drains the Filter Ring and, for each dequeued flow entry, constructs a 5-tuple filter rule and installs it on the NIC hardware through the DPDK \texttt{rte\_flow} interface. The rule matches the flow's 5-tuple and applies an ingress drop action to subsequent matched packets.}
Unlike a synchronous design where rule update would block the model inference path, the asynchronous engine fully decouples NIC hardware communication from model inference, ensuring that the latency of rule update does not affect model analysis latency. 
Combined with the parallel launch from the packet processing engine, this design minimizes the time window between the first packet of a flow and the moment its subsequent packets are filtered at line rate on the NIC.

%% file: Evaluation.tex
\section{Experiments}
\label{sec:Experiments}

We prototype \textit{Nepco} in both software and hardware, and evaluate it extensively on six public network traffic datasets, including 83 different traffic categories, against 17 baselines covering traditional ML/DL models, hardware-offloadable models, and network foundation models.
In general, the experiments demonstrate that \textit{Nepco} is able to:

\begin{enumerate}
    \item \cg{achieve competitive classification performance with state-of-the-art network foundation models under limited labeled data, while showing significantly lower label dependency than existing hardware-offloadable models under the constrained early-stage analysis setting (Section~\ref{sec:Versatility Evaluation}).}
    
    
    \item \cg{reduce the analysis latency of network foundation models from second-scale to millisecond-scale under the SmartNIC deployment setting, while preserving nanosecond-scale input construction and microsecond-scale queue wait and filter rule installation latencies (Section~\ref{sec:Efficiency Evaluation}).}
    
    \item validate the effectiveness of the localized pattern-aware traffic modeling over traffic byte sequences, achieving simultaneous improvements in both classification performance and analysis latency over existing convolution-based methods (Section~\ref{sec:Nepco Deep Dive}).
\end{enumerate}

\subsection{Experiment Setup}

\noindent\textbf{Implementation.} We develop both software and hardware prototypes of \textit{Nepco} \cg{\footnote{\cg{https://github.com/lincgcg//Nepco}}}.
The software prototype follows the same deployment setting as existing network foundation models, running on an end-host server equipped with an NVIDIA H800 GPU \cite{H800} and an Intel Xeon Platinum 8458P CPU \cite{Xeon8458P}.
We implement the traffic modeling component using PyTorch (v2.3.0 for CUDA v12.4).
The hardware prototype is implemented on an NVIDIA BlueField-3 SmartNIC \cite{BF3}, whose NIC is based on ConnectX-7 \cite{CX7}.
We leverage the programmability of the NIC hardware to implement the NIC enforcement engine. 
The SoC is equipped with 16 cores, which we allocate across the four engines to maximize throughput while ensuring real-time responsiveness.
Specifically, 13 cores are dedicated to the traffic analysis engine to parallelize input construction and model inference across concurrent flows, 1 core is assigned to the packet processing engine, and 1 core to the filter-rule management engine.
The remaining core is reserved to coordinate inter-engine communication and prevent resource exhaustion.
For both prototypes of \textit{Nepco}, we configure the traffic modeling component with a byte embedding size of 128, a kernel size of 4 in the traffic convolution layer, and a stack of 5 consecutive traffic convolution layers.

\begin{table}[t]
\renewcommand{\arraystretch}{0.85}
\centering
\caption{Datasets for pre-training and fine-tuning.}
\label{tab:dataset-summary}
\resizebox{\columnwidth}{!}{%
\begin{tabular}{c|c|c|c|c}
\toprule
\textbf{Dataset} &
\textbf{Year} &
\textbf{Flows} &
\textbf{Pre-train} &
\textbf{Fine-tune} \\ \midrule
Bottlenet-HTTPS \cite{Bottlenet-HTTPS-Website} & 2016 & 610,161 & \ding{51} & \ding{55} \\ \midrule

ISCX-SlowDoS \cite{ISCX-SlowDos-CN24} & 2016 & 176,211 & \ding{51} & \ding{55} \\ \midrule

USTC-TFC \cite{USTC-TFC-ICOIN17} & 2016 & 179,252 & \ding{55} & \makecell[c]{\ding{51} 10-class\\ Malware Family Classification} \\ \midrule

ISCX-VPN-App \cite{ISCX-VPN-ICISSP16} & 2016 & 292,922 & \ding{55} & \makecell[c]{\ding{51} 17-class\\ Application Classification} \\ \midrule

ISCX-VPN-Service \cite{ISCX-VPN-ICISSP16} & 2016 & 101,354 & \ding{55} & \makecell[c]{\ding{51} 12-class\\ Service Type Identification} \\ \midrule

DataCon \cite{DataCon2020EncryptedMaliciousTrafficDataset} & 2021 & 60,509 & \ding{55} & \makecell[c]{\ding{51} 10-class\\ Encrypted Proxy Identification} \\ \midrule

CIC-EVSE \cite{CIC-EVSE-DASP24} & 2024 & 2,743,495 & \ding{55} & \makecell[c]{\ding{51} 15-class\\ EVSE Intrusion Classification} \\ \midrule

CIC-IoMT \cite{CIC-IoMT-arXiv24} & 2024 & 6,998,544 & \ding{55} & \makecell[c]{\ding{51} 19-class\\ IoMT Intrusion Classification} \\
\bottomrule

\end{tabular}%
}
\vspace{-8pt}
\end{table}

\noindent\textbf{Testbed.}
We use a testbed with two servers directly connected with a $200$Gbps link, each equipped with a $5.4$GHz Intel Core i7-14700 CPU, $64$GB DRAM, and an NVIDIA BlueField-3 SmartNIC.
We generate traffic in one server (sender) and deploy {\it Nepco} on the other server (receiver).
We implement two traffic generation methods targeting different evaluation scenarios.
For real and continuous latency evaluation, we use DPDK to generate traffic based on packet length and inter-packet interval distributions collected from backbone networks, faithfully reflecting real-world network conditions.
We acknowledge that direct deployment in a production environment is infeasible due to privacy and access constraints; the use of distribution-matched synthetic traffic is a standard methodology in prior work \cite{Exosphere-CCS24,SmartTC-IWQoS25,N3ID-NSDI22}.
For latency evaluation under diverse network load, we use TCP raw sockets \cite{LinuxRawSocket} to manually construct IP/TCP headers under controlled New Flow Per Second (NFPS) settings.
We estimate a realistic baseline load based on prior measurements. Meta~\cite{Meta_NFPS} reported in 2015 that its external-facing web servers observe 500 NFPS at the median, and Cisco~\cite{Cisco_NFPS} measured a 3$\times$ growth in Internet traffic from 2016 to 2021, suggesting 2,000 NFPS as a reasonable network load for practical deployments.
Accordingly, we consider four network load settings, i.e., Extreme Low (NFPS~=~5), Low (NFPS~=~500), Normal (NFPS~=~2,000), and High (NFPS~=~5,000), during our evaluation.

\noindent\textbf{Pre-training.}
We pre-train \textit{Nepco} on two public network traffic datasets, i.e., Bottlenet-HTTPS~\cite{Bottlenet-HTTPS-Website} and ISCX-SlowDoS~\cite{ISCX-SlowDos-CN24}, comprising 786,372 flows in total, as summarized in TABLE~\ref{tab:dataset-summary}.
Both datasets were collected in 2016 and are kept strictly disjoint from all fine-tuning datasets to avoid biased estimates of generalization, following the evaluation safeguards described in Section~\ref{sec:Versatile Analysis with Foundation Models}.
The pre-training runs for 100,000 steps with a learning rate of $1 \times 10^{-3}$.

\noindent\textbf{Fine-tuning.}
We fine-tune and evaluate \textit{Nepco} spanning three security analysis tasks, including malware analysis, application fingerprinting, and intrusion detection, as shown in TABLE~\ref{tab:dataset-summary}. For malware analysis, we evaluate on USTC-TFC~\cite{USTC-TFC-ICOIN17} to perform fine-grained malware family classification across 10 classes. For application fingerprinting, we evaluate on ISCX-VPN-App~\cite{ISCX-VPN-ICISSP16}, ISCX-VPN-Service~\cite{ISCX-VPN-ICISSP16}, and DataCon~\cite{DataCon2020EncryptedMaliciousTrafficDataset} to cover three distinct fingerprinting scenarios, i.e., application classification, service type identification, and encrypted proxy identification, comprising 39 classes in total.
For intrusion detection, we evaluate on CIC-EVSE~\cite{CIC-EVSE-DASP24} and CIC-IoMT~\cite{CIC-IoMT-arXiv24} to perform fine-grained intrusion classification across two emerging IoT deployment scenarios, i.e., EV charging stations and IoMT devices.
For each fine-tuning dataset, we split the data into training, validation, and test sets at a ratio of 8:1:1, and construct 20 sampled subsets using different random seeds for statistical significance analysis.
The fine-tuning runs for 10 epochs with a learning rate of $5 \times 10^{-4}$.

\noindent\textbf{Baselines.} We compare \textit{Nepco} with 17 baselines, including traditional ML/DL models, hardware-offloadable models, and network foundation models, as shown in TABLE~\ref{tab:Overall Classification Performance}.
\cg{To measure the analysis latency of existing network foundation models on SmartNICs, we deploy three network foundation models (i.e., ET-BERT, TraGe, and TrafficFormer) onto the same hardware prototype under \textit{Nepco}'s deployment scheme, with only the traffic processing and model architecture varying across baselines.}
For existing hardware-offloadable models, since most are designed for FPGAs \cite{N3ID-NSDI22,Exosphere-CCS24} or P4 switches \cite{Flowrest-INFOCOM23,Leo-NSDI24,Mousika-ToN23,BoS-NSDI24}, we implement their software prototypes with corresponding open-source codes to evaluate versatility on the host CPU.
Since \textit{Nepco} is a convolution-based network foundation model, we further consider 4 traditional convolution-based methods~\cite{Text-CNN-EMNLP14,VGG-arXiv14,GoogLeNet-CVPR15,LeNeet-5-IEEE02} 
to validate the effectiveness of \textit{Nepco} design.

\begin{table*}[t!]
\renewcommand{\arraystretch}{1.1}
\centering
\small
\caption{The overall classification performances of \textit{Nepco} and baselines on six public network traffic datasets.}
\label{tab:Overall Classification Performance}
\resizebox{1\textwidth}{!}{%
\begin{tabular}{c|c|ccc|ccc|ccc}

\toprule
\multicolumn{2}{c|}{\multirow{2}{*}{\textbf{Method} \cg{\textsuperscript{1}}}} &
\multicolumn{3}{c|}{\textbf{DataCon (2021)}} &
\multicolumn{3}{c|}{\textbf{CIC-EVSE (2024)}} &
\multicolumn{3}{c}{\textbf{CIC-IoMT (2024)}} \\

\multicolumn{2}{c|}{} &
Precision &
Recall &
Macro F1 &
Precision &
Recall &
Macro F1 &
Precision &
Recall &
Macro F1 \\ \midrule

\multirow{9}{*}{\rotatebox[origin=c]{90}{\makecell{ML/DL Model \\ \textbf{\textcolor{CheckBlue}{\ding{51}}}: Hardware-Offloadable}}} &
FlowPrint &
0.0169 {\scriptsize\(\pm \ 0.0001\)} &
0.1000 {\scriptsize\(\pm \ 0.0000\)} &
0.0289 {\scriptsize\(\pm \ 0.0002\)} &
0.2379 {\scriptsize\(\pm \ 0.0461\)} &
0.1890 {\scriptsize\(\pm \ 0.0270\)} &
0.1616 {\scriptsize\(\pm \ 0.0256\)} &
0.3625 {\scriptsize\(\pm \ 0.0311\)} &
0.3116 {\scriptsize\(\pm \ 0.0209\)} &
0.2827 {\scriptsize\(\pm \ 0.0230\)} \\

 &
AppScanner &
0.6979 {\scriptsize\(\pm \ 0.0112\)} &
0.6428 {\scriptsize\(\pm \ 0.0077\)} &
0.6454 {\scriptsize\(\pm \ 0.0078\)} &
0.4881 {\scriptsize\(\pm \ 0.0203\)} &
0.3042 {\scriptsize\(\pm \ 0.0119\)} &
0.2838 {\scriptsize\(\pm \ 0.0100\)} &
0.5091 {\scriptsize\(\pm \ 0.0100\)} &
0.2483 {\scriptsize\(\pm \ 0.0012\)} &
0.2450 {\scriptsize\(\pm \ 0.0018\)} \\

 &
XGBoost &
0.8482 {\scriptsize\(\pm \ 0.0063\)} &
0.8087 {\scriptsize\(\pm \ 0.0081\)} &
0.8208 {\scriptsize\(\pm \ 0.0057\)} &
0.6412 {\scriptsize\(\pm \ 0.0078\)} &
0.6290 {\scriptsize\(\pm \ 0.0054\)} &
0.6188 {\scriptsize\(\pm \ 0.0073\)} &
0.4512 {\scriptsize\(\pm \ 0.0263\)} &
0.4315 {\scriptsize\(\pm \ 0.0122\)} &
0.4410 {\scriptsize\(\pm \ 0.0193\)} \\

 &
DTree &
0.8130 {\scriptsize\(\pm \ 0.0076\)} &
0.8123 {\scriptsize\(\pm \ 0.0090\)} &
0.8080 {\scriptsize\(\pm \ 0.0054\)} &
0.6313 {\scriptsize\(\pm \ 0.0108\)} &
0.6215 {\scriptsize\(\pm \ 0.0102\)} &
0.6181 {\scriptsize\(\pm \ 0.0084\)} &
0.4135 {\scriptsize\(\pm \ 0.0123\)} &
0.4211 {\scriptsize\(\pm \ 0.0106\)} &
0.4122 {\scriptsize\(\pm \ 0.0102\)} \\

 &
Flowrest \textbf{\textcolor{CheckBlue}{\ding{51}}} &
0.4187 {\scriptsize\(\pm \ 0.0200\)} &
0.3906 {\scriptsize\(\pm \ 0.0052\)} &
0.3891 {\scriptsize\(\pm \ 0.0067\)} &
\textbf{\textcolor{CheckBlue}{\uline{0.5839 {\scriptsize\(\pm \ 0.0251\)}}}} &
\textbf{\textcolor{CheckBlue}{\uline{0.5815 {\scriptsize\(\pm \ 0.0167\)}}}} &
\textbf{\textcolor{CheckBlue}{\uline{0.5424 {\scriptsize\(\pm \ 0.0176\)}}}} &
0.4372 {\scriptsize\(\pm \ 0.0188\)} &
0.3914 {\scriptsize\(\pm \ 0.0051\)} &
0.3767 {\scriptsize\(\pm \ 0.0068\)} \\

 &
Leo \textbf{\textcolor{CheckBlue}{\ding{51}}} &
0.3976 {\scriptsize\(\pm \ 0.0225\)} &
0.5981 {\scriptsize\(\pm \ 0.0057\)} &
0.3720 {\scriptsize\(\pm \ 0.0194\)} &
0.3718 {\scriptsize\(\pm \ 0.0218\)} &
0.4929 {\scriptsize\(\pm \ 0.0135\)} &
0.3604 {\scriptsize\(\pm \ 0.0127\)} &
\textbf{\textcolor{CheckBlue}{\uline{0.4917 {\scriptsize\(\pm \ 0.0077\)}}}} &
\textbf{\textcolor{CheckBlue}{\uline{0.4964 {\scriptsize\(\pm \ 0.0034\)}}}} &
\textbf{\textcolor{CheckBlue}{\uline{0.4398 {\scriptsize\(\pm \ 0.0028\)}}}} \\

 &
Mousika \textbf{\textcolor{CheckBlue}{\ding{51}}} &
\textbf{\textcolor{CheckBlue}{\uline{0.8642 {\scriptsize\(\pm \ 0.0025\)}}}} &
\textbf{\textcolor{CheckBlue}{\uline{0.7892 {\scriptsize\(\pm \ 0.0059\)}}}} &
\textbf{\textcolor{CheckBlue}{\uline{0.8008 {\scriptsize\(\pm \ 0.0041\)}}}} &
0.3435 {\scriptsize\(\pm \ 0.0229\)} &
0.3614 {\scriptsize\(\pm \ 0.0173\)} &
0.3181 {\scriptsize\(\pm \ 0.0178\)} &
0.4518 {\scriptsize\(\pm \ 0.0110\)} &
0.3925 {\scriptsize\(\pm \ 0.0037\)} &
0.3497 {\scriptsize\(\pm \ 0.0042\)} \\

 &
EBSNN &
0.9448 {\scriptsize\(\pm \ 0.0195\)} &
0.9695 {\scriptsize\(\pm \ 0.0097\)} &
0.9501 {\scriptsize\(\pm \ 0.0199\)} &
0.8238 {\scriptsize\(\pm \ 0.0286\)} &
0.8267 {\scriptsize\(\pm \ 0.0208\)} &
0.8028 {\scriptsize\(\pm \ 0.0210\)} &
0.8289 {\scriptsize\(\pm \ 0.0139\)} &
0.7760 {\scriptsize\(\pm \ 0.0074\)} &
0.7671 {\scriptsize\(\pm \ 0.0077\)} \\

 &
BoS \textbf{\textcolor{CheckBlue}{\ding{51}}} &
0.3792 {\scriptsize\(\pm \ 0.0219\)} &
0.3554 {\scriptsize\(\pm \ 0.0216\)} &
0.3055 {\scriptsize\(\pm \ 0.0089\)} &
0.1392 {\scriptsize\(\pm \ 0.0086\)} &
0.2322 {\scriptsize\(\pm \ 0.0135\)} &
0.1378 {\scriptsize\(\pm \ 0.0070\)} &
0.1249 {\scriptsize\(\pm \ 0.0050\)} &
0.2382 {\scriptsize\(\pm \ 0.0033\)} &
0.1434 {\scriptsize\(\pm \ 0.0024\)} \\

\midrule

\multirow{9}{*}{\rotatebox[origin=c]{90}{\makecell{Network Foundation Model \\ \textbf{\textcolor{CheckBlue}{\ding{51}}}: Hardware-Offloadable}}} &
NetMamba &
0.9871 {\scriptsize\(\pm \ 0.0050\)} &
0.9875 {\scriptsize\(\pm \ 0.0076\)} &
0.9870 {\scriptsize\(\pm \ 0.0063\)} &
0.8138 {\scriptsize\(\pm \ 0.0099\)} &
0.7666 {\scriptsize\(\pm \ 0.0060\)} &
0.7601 {\scriptsize\(\pm \ 0.0062\)} &
0.4728 {\scriptsize\(\pm \ 0.1641\)} &
0.4645 {\scriptsize\(\pm \ 0.1552\)} &
0.4463 {\scriptsize\(\pm \ 0.1541\)} \\

 &
YaTC &
0.9871 {\scriptsize\(\pm \ 0.0026\)} &
0.9793 {\scriptsize\(\pm \ 0.0066\)} &
0.9821 {\scriptsize\(\pm \ 0.0042\)} &
0.8120 {\scriptsize\(\pm \ 0.0047\)} &
0.7531 {\scriptsize\(\pm \ 0.0100\)} &
0.7511 {\scriptsize\(\pm \ 0.0089\)} &
0.7520 {\scriptsize\(\pm \ 0.0187\)} &
0.7171 {\scriptsize\(\pm \ 0.0151\)} &
0.6990 {\scriptsize\(\pm \ 0.0168\)} \\

 &
Pcap-Encoder &
0.5855 {\scriptsize\(\pm \ 0.0047\)} &
0.5947 {\scriptsize\(\pm \ 0.0089\)} &
0.5896 {\scriptsize\(\pm \ 0.0060\)} &
0.4302 {\scriptsize\(\pm \ 0.0339\)} &
0.4476 {\scriptsize\(\pm \ 0.0282\)} &
0.4051 {\scriptsize\(\pm \ 0.0320\)} &
0.5344 {\scriptsize\(\pm \ 0.0139\)} &
0.5632 {\scriptsize\(\pm \ 0.0057\)} &
0.5308 {\scriptsize\(\pm \ 0.0092\)} \\

 &
PERT &
\textbf{\textcolor{CrossRed}{0.9927 {\scriptsize\(\pm \ 0.0018\)}}} &
0.9888 {\scriptsize\(\pm \ 0.0049\)} &
0.9902 {\scriptsize\(\pm \ 0.0033\)} &
0.8588 {\scriptsize\(\pm \ 0.0247\)} &
0.8496 {\scriptsize\(\pm \ 0.0233\)} &
0.8463 {\scriptsize\(\pm \ 0.0231\)} &
0.8210 {\scriptsize\(\pm \ 0.0045\)} &
0.8175 {\scriptsize\(\pm \ 0.0069\)} &
0.8144 {\scriptsize\(\pm \ 0.0055\)} \\

 &
NetGPT &
0.9914 {\scriptsize\(\pm \ 0.0016\)} &
0.9872 {\scriptsize\(\pm \ 0.0032\)} &
0.9890 {\scriptsize\(\pm \ 0.0020\)} &
\textbf{\textcolor{CrossRed}{0.8622 {\scriptsize\(\pm \ 0.0245\)}}} &
\textbf{\textcolor{CrossRed}{0.8712 {\scriptsize\(\pm \ 0.0217\)}}} &
\textbf{\textcolor{CrossRed}{0.8588 {\scriptsize\(\pm \ 0.0224\)}}} &
0.8267 {\scriptsize\(\pm \ 0.0067\)} &
0.8104 {\scriptsize\(\pm \ 0.0088\)} &
0.8076 {\scriptsize\(\pm \ 0.0098\)} \\

 &
ET-BERT &
0.9920 {\scriptsize\(\pm \ 0.0021\)} &
\textbf{\textcolor{CrossRed}{0.9900 {\scriptsize\(\pm \ 0.0024\)}}} &
\textbf{\textcolor{CrossRed}{0.9908 {\scriptsize\(\pm \ 0.0017\)}}} &
0.8594 {\scriptsize\(\pm \ 0.0203\)} &
0.8631 {\scriptsize\(\pm \ 0.0208\)} &
0.8519 {\scriptsize\(\pm \ 0.0198\)} &
\textbf{\textcolor{CrossRed}{0.8373 {\scriptsize\(\pm \ 0.0070\)}}} &
0.8223 {\scriptsize\(\pm \ 0.0082\)} &
0.8187 {\scriptsize\(\pm \ 0.0077\)} \\

 &
TraGe &
0.9911 {\scriptsize\(\pm \ 0.0025\)} &
0.9885 {\scriptsize\(\pm \ 0.0021\)} &
0.9881 {\scriptsize\(\pm \ 0.0023\)} &
0.8343 {\scriptsize\(\pm \ 0.0114\)} &
0.8219 {\scriptsize\(\pm \ 0.0187\)} &
0.8220 {\scriptsize\(\pm \ 0.0191\)} &
0.8254 {\scriptsize\(\pm \ 0.0077\)} &
0.8174 {\scriptsize\(\pm \ 0.0035\)} &
0.8177 {\scriptsize\(\pm \ 0.0066\)} \\

 &
TrafficFormer &
0.9914 {\scriptsize\(\pm \ 0.0031\)} &
0.9855 {\scriptsize\(\pm \ 0.0045\)} &
0.9881 {\scriptsize\(\pm \ 0.0034\)} &
0.8555 {\scriptsize\(\pm \ 0.0196\)} &
0.8580 {\scriptsize\(\pm \ 0.0195\)} &
0.8483 {\scriptsize\(\pm \ 0.0194\)} &
0.8313 {\scriptsize\(\pm \ 0.0055\)} &
\textbf{\textcolor{CrossRed}{0.8261 {\scriptsize\(\pm \ 0.0056\)}}} &
\textbf{\textcolor{CrossRed}{0.8235 {\scriptsize\(\pm \ 0.0053\)}}} \\  \cmidrule(lr){2-11}

 &
\textit{\textbf{Nepco \textbf{\textcolor{CheckBlue}{\ding{51}}}}} &
\textbf{\uline{0.9921 {\scriptsize\(\pm \ 0.0009\)}}} &
\textbf{\uline{0.9909 {\scriptsize\(\pm \ 0.0015\)}}} &
\textbf{\uline{0.9914 {\scriptsize\(\pm \ 0.0009\)}}} &
\textbf{\uline{0.8922 {\scriptsize\(\pm \ 0.0159\)}}} &
\textbf{\uline{0.8899 {\scriptsize\(\pm \ 0.0128\)}}} &
\textbf{\uline{0.8838 {\scriptsize\(\pm \ 0.0122\)}}} &
\textbf{\uline{0.8389 {\scriptsize\(\pm \ 0.0041\)}}} &
\textbf{\uline{0.8291 {\scriptsize\(\pm \ 0.0061\)}}} &
\textbf{\uline{0.8341 {\scriptsize\(\pm \ 0.0052\)}}} \\ 

\bottomrule

\addlinespace[4pt]

\toprule

\multicolumn{2}{c|}{\multirow{2}{*}{\textbf{Method}}} &
\multicolumn{3}{c|}{\textbf{ISCX-VPN-App (2016)}} &
\multicolumn{3}{c|}{\textbf{ISCX-VPN-Service (2016)}} &
\multicolumn{3}{c}{\textbf{USTC-TFC (2016)}} \\

\multicolumn{2}{c|}{} &

Precision &
Recall &
Macro F1 &
Precision &
Recall &
Macro F1 &
Precision &
Recall &
Macro F1 \\ \midrule

\multirow{9}{*}{\rotatebox[origin=c]{90}{\makecell{ML/DL Model \\ \textbf{\textcolor{CheckBlue}{\ding{51}}}: Hardware-Offloadable}}} &
FlowPrint &
0.5959 {\scriptsize\(\pm \ 0.0139\)} &
0.4115 {\scriptsize\(\pm \ 0.0112\)} &
0.4334 {\scriptsize\(\pm \ 0.0140\)} &
0.7166 {\scriptsize\(\pm \ 0.0215\)} &
0.6860 {\scriptsize\(\pm \ 0.0187\)} &
0.6638 {\scriptsize\(\pm \ 0.0181\)} &
0.6764 {\scriptsize\(\pm \ 0.0158\)} &
0.6693 {\scriptsize\(\pm \ 0.0135\)} &
0.6636 {\scriptsize\(\pm \ 0.0141\)} \\

 &
AppScanner &
0.7394 {\scriptsize\(\pm \ 0.0096\)} &
0.5439 {\scriptsize\(\pm \ 0.0098\)} &
0.5873 {\scriptsize\(\pm \ 0.0099\)} &
0.8567 {\scriptsize\(\pm \ 0.0079\)} &
0.7586 {\scriptsize\(\pm \ 0.0080\)} &
0.7897 {\scriptsize\(\pm \ 0.0071\)} &
0.8377 {\scriptsize\(\pm \ 0.0049\)} &
0.6408 {\scriptsize\(\pm \ 0.0044\)} &
0.6966 {\scriptsize\(\pm \ 0.0043\)} \\

 &
XGBoost &
0.5177 {\scriptsize\(\pm \ 0.0104\)} &
0.4151 {\scriptsize\(\pm \ 0.0050\)} &
0.4304 {\scriptsize\(\pm \ 0.0057\)} &
0.7094 {\scriptsize\(\pm \ 0.0033\)} &
0.7160 {\scriptsize\(\pm \ 0.0036\)} &
0.7032 {\scriptsize\(\pm \ 0.0029\)} &
0.9438 {\scriptsize\(\pm \ 0.0013\)} &
0.9435 {\scriptsize\(\pm \ 0.0014\)} &
0.9434 {\scriptsize\(\pm \ 0.0014\)} \\

 &
DTree &
0.4844 {\scriptsize\(\pm \ 0.0056\)} &
0.4483 {\scriptsize\(\pm \ 0.0050\)} &
0.4462 {\scriptsize\(\pm \ 0.0047\)} &
0.7026 {\scriptsize\(\pm \ 0.0030\)} &
0.7090 {\scriptsize\(\pm \ 0.0027\)} &
0.6997 {\scriptsize\(\pm \ 0.0023\)} &
0.9411 {\scriptsize\(\pm \ 0.0010\)} &
0.9299 {\scriptsize\(\pm \ 0.0035\)} &
0.9280 {\scriptsize\(\pm \ 0.0040\)} \\

 &
Flowrest \textbf{\textcolor{CheckBlue}{\ding{51}}} &
0.2983 {\scriptsize\(\pm \ 0.0136\)} &
0.2350 {\scriptsize\(\pm \ 0.0050\)} &
0.2278 {\scriptsize\(\pm \ 0.0061\)} &
0.5263 {\scriptsize\(\pm \ 0.0205\)} &
0.4572 {\scriptsize\(\pm \ 0.0100\)} &
0.4429 {\scriptsize\(\pm \ 0.0111\)} &
0.7015 {\scriptsize\(\pm \ 0.0092\)} &
0.7049 {\scriptsize\(\pm \ 0.0084\)} &
0.6869 {\scriptsize\(\pm \ 0.0078\)} \\

 &
Leo \textbf{\textcolor{CheckBlue}{\ding{51}}} &
0.1839 {\scriptsize\(\pm \ 0.0083\)} &
0.2645 {\scriptsize\(\pm \ 0.0043\)} &
0.1543 {\scriptsize\(\pm \ 0.0029\)} &
0.4105 {\scriptsize\(\pm \ 0.0048\)} &
\textbf{\textcolor{CheckBlue}{\uline{0.5097 {\scriptsize\(\pm \ 0.0065\)}}}} &
0.4193 {\scriptsize\(\pm \ 0.0057\)} &
\textbf{\textcolor{CheckBlue}{\uline{0.8268 {\scriptsize\(\pm \ 0.0035\)}}}} &
\textbf{\textcolor{CheckBlue}{\uline{0.8130 {\scriptsize\(\pm \ 0.0028\)}}}} &
\textbf{\textcolor{CheckBlue}{\uline{0.8104 {\scriptsize\(\pm \ 0.0026\)}}}} \\

 &
Mousika \textbf{\textcolor{CheckBlue}{\ding{51}}} &
\textbf{\textcolor{CheckBlue}{\uline{0.4849 {\scriptsize\(\pm \ 0.0154\)}}}} &
\textbf{\textcolor{CheckBlue}{\uline{0.3221 {\scriptsize\(\pm \ 0.0061\)}}}} &
\textbf{\textcolor{CheckBlue}{\uline{0.3370 {\scriptsize\(\pm \ 0.0070\)}}}} &
\textbf{\textcolor{CheckBlue}{\uline{0.6336 {\scriptsize\(\pm \ 0.0179\)}}}} &
0.4523 {\scriptsize\(\pm \ 0.0058\)} &
\textbf{\textcolor{CheckBlue}{\uline{0.4545 {\scriptsize\(\pm \ 0.0070\)}}}} &
0.7291 {\scriptsize\(\pm \ 0.0111\)} &
0.7118 {\scriptsize\(\pm \ 0.0013\)} &
0.6615 {\scriptsize\(\pm \ 0.0018\)} \\

 &
EBSNN &
0.7154 {\scriptsize\(\pm \ 0.0107\)} &
0.6560 {\scriptsize\(\pm \ 0.0081\)} &
0.6620 {\scriptsize\(\pm \ 0.0086\)} &
0.8945 {\scriptsize\(\pm \ 0.0054\)} &
0.8757 {\scriptsize\(\pm \ 0.0070\)} &
0.8804 {\scriptsize\(\pm \ 0.0054\)} &
0.9020 {\scriptsize\(\pm \ 0.0071\)} &
0.8780 {\scriptsize\(\pm \ 0.0120\)} &
0.8667 {\scriptsize\(\pm \ 0.0156\)} \\

 &
BoS \textbf{\textcolor{CheckBlue}{\ding{51}}} &
0.2473 {\scriptsize\(\pm \ 0.0140\)} &
0.2549 {\scriptsize\(\pm \ 0.0069\)} &
0.2161 {\scriptsize\(\pm \ 0.0061\)} &
0.4865 {\scriptsize\(\pm \ 0.0225\)} &
0.4594 {\scriptsize\(\pm \ 0.0170\)} &
0.4230 {\scriptsize\(\pm \ 0.0181\)} &
0.6070 {\scriptsize\(\pm \ 0.0196\)} &
0.6245 {\scriptsize\(\pm \ 0.0128\)} &
0.5737 {\scriptsize\(\pm \ 0.0141\)} \\

\midrule

\multirow{9}{*}{\rotatebox[origin=c]{90}{\makecell{Network Foundation Model \\ \textbf{\textcolor{CheckBlue}{\ding{51}}}: Hardware-Offloadable}}} &
NetMamba &
0.6834 {\scriptsize\(\pm \ 0.0163\)} &
0.5984 {\scriptsize\(\pm \ 0.0110\)} &
0.6164 {\scriptsize\(\pm \ 0.0122\)} &
0.8421 {\scriptsize\(\pm \ 0.0045\)} &
0.7950 {\scriptsize\(\pm \ 0.0040\)} &
0.8030 {\scriptsize\(\pm \ 0.0033\)} &
\textbf{\textcolor{CrossRed}{0.9655 {\scriptsize\(\pm \ 0.0015\)}}} &
\textbf{\textcolor{CrossRed}{0.9631 {\scriptsize\(\pm \ 0.0012\)}}} &
\textbf{\textcolor{CrossRed}{0.9628 {\scriptsize\(\pm \ 0.0013\)}}} \\

 &
YaTC &
0.6531 {\scriptsize\(\pm \ 0.0166\)} &
0.5764 {\scriptsize\(\pm \ 0.0097\)} &
0.5845 {\scriptsize\(\pm \ 0.0116\)} &
0.8331 {\scriptsize\(\pm \ 0.0045\)} &
0.7854 {\scriptsize\(\pm \ 0.0042\)} &
0.7937 {\scriptsize\(\pm \ 0.0031\)} &
0.9644 {\scriptsize\(\pm \ 0.0009\)} &
0.9605 {\scriptsize\(\pm \ 0.0015\)} &
0.9600 {\scriptsize\(\pm \ 0.0017\)} \\

 &
Pcap-Encoder &
0.2951 {\scriptsize\(\pm \ 0.0312\)} &
0.3060 {\scriptsize\(\pm \ 0.0231\)} &
0.2813 {\scriptsize\(\pm \ 0.0255\)} &
0.5357 {\scriptsize\(\pm \ 0.0504\)} &
0.5203 {\scriptsize\(\pm \ 0.0355\)} &
0.5020 {\scriptsize\(\pm \ 0.0397\)} &
0.9298 {\scriptsize\(\pm \ 0.0090\)} &
0.9232 {\scriptsize\(\pm \ 0.0105\)} &
0.9228 {\scriptsize\(\pm \ 0.0108\)} \\

 &
PERT &
0.7166 {\scriptsize\(\pm \ 0.0089\)} &
0.7004 {\scriptsize\(\pm \ 0.0073\)} &
0.7003 {\scriptsize\(\pm \ 0.0067\)} &
0.9146 {\scriptsize\(\pm \ 0.0033\)} &
0.9104 {\scriptsize\(\pm \ 0.0020\)} &
0.9115 {\scriptsize\(\pm \ 0.0020\)} &
0.9345 {\scriptsize\(\pm \ 0.0022\)} &
0.9337 {\scriptsize\(\pm \ 0.0023\)} &
0.9337 {\scriptsize\(\pm \ 0.0023\)} \\

 &
NetGPT &
0.7222 {\scriptsize\(\pm \ 0.0065\)} &
\textbf{\textcolor{CrossRed}{0.7226 {\scriptsize\(\pm \ 0.0097\)}}} &
\textbf{\textcolor{CrossRed}{0.7154 {\scriptsize\(\pm \ 0.0091\)}}} &
0.9211 {\scriptsize\(\pm \ 0.0028\)} &
0.9196 {\scriptsize\(\pm \ 0.0019\)} &
0.9190 {\scriptsize\(\pm \ 0.0014\)} &
0.9607 {\scriptsize\(\pm \ 0.0017\)} &
0.9597 {\scriptsize\(\pm \ 0.0017\)} &
0.9596 {\scriptsize\(\pm \ 0.0017\)} \\

 &
ET-BERT &
\textbf{\textcolor{CrossRed}{0.7352 {\scriptsize\(\pm \ 0.0096\)}}} &
0.7103 {\scriptsize\(\pm \ 0.0092\)} &
0.7125 {\scriptsize\(\pm \ 0.0081\)} &
0.9225 {\scriptsize\(\pm \ 0.0025\)} &
0.9236 {\scriptsize\(\pm \ 0.0023\)} &
0.9224 {\scriptsize\(\pm \ 0.0020\)} &
0.9525 {\scriptsize\(\pm \ 0.0019\)} &
0.9511 {\scriptsize\(\pm \ 0.0020\)} &
0.9509 {\scriptsize\(\pm \ 0.0020\)} \\

 &
TraGe &
0.7212 {\scriptsize\(\pm \ 0.0122\)} &
0.7132 {\scriptsize\(\pm \ 0.0069\)} &
0.7122 {\scriptsize\(\pm \ 0.0098\)} &
0.9202 {\scriptsize\(\pm \ 0.0022\)} &
0.9193 {\scriptsize\(\pm \ 0.0025\)} &
0.9191 {\scriptsize\(\pm \ 0.0022\)} &
0.9534 {\scriptsize\(\pm \ 0.0018\)} &
0.9524 {\scriptsize\(\pm \ 0.0032\)} &
0.9514 {\scriptsize\(\pm \ 0.0023\)} \\

 &
TrafficFormer &
0.7238 {\scriptsize\(\pm \ 0.0081\)} &
0.7110 {\scriptsize\(\pm \ 0.0088\)} &
0.7111 {\scriptsize\(\pm \ 0.0088\)} &
\textbf{\textcolor{CrossRed}{0.9238 {\scriptsize\(\pm \ 0.0025\)}}} &
\textbf{\textcolor{CrossRed}{0.9237 {\scriptsize\(\pm \ 0.0021\)}}} &
\textbf{\textcolor{CrossRed}{0.9230 {\scriptsize\(\pm \ 0.0017\)}}} &
0.9489 {\scriptsize\(\pm \ 0.0026\)} &
0.9471 {\scriptsize\(\pm \ 0.0025\)} &
0.9469 {\scriptsize\(\pm \ 0.0025\)} \\ \cmidrule(lr){2-11}

 &
\textit{\textbf{Nepco \textbf{\textcolor{CheckBlue}{\ding{51}}}}} &
\textbf{\uline{0.7032 {\scriptsize\(\pm \ 0.0079\)}}} &
\textbf{\uline{0.7071 {\scriptsize\(\pm \ 0.0068\)}}} &
\textbf{\uline{0.7034 {\scriptsize\(\pm \ 0.0078\)}}} &
\textbf{\uline{0.9248 {\scriptsize\(\pm \ 0.0024\)}}} &
\textbf{\uline{0.9190 {\scriptsize\(\pm \ 0.0028\)}}} &
\textbf{\uline{0.9209 {\scriptsize\(\pm \ 0.0022\)}}} &
\textbf{\uline{0.9643 {\scriptsize\(\pm \ 0.0021\)}}} &
\textbf{\uline{0.9636 {\scriptsize\(\pm \ 0.0019\)}}} &
\textbf{\uline{0.9639 {\scriptsize\(\pm \ 0.0021\)}}} \\ 

\bottomrule

\multicolumn{11}{l}{\cg{\textsuperscript{1} We consider the early-stage analysis setting using the first 5 packets of each flow, which represents a constrained-setting comparison for hardware-offloadable models with longer observations.}} \\

\multicolumn{11}{l}{\cg{\textsuperscript{2} We further report the overall classification performance of \textit{Nepco} and baselines on three additional network traffic datasets (i.e., CSTNET-20, CSTNET-120, and DoHBrw) in Appendix \ref{app:Supporting Materials}.}} \\

\end{tabular}%
}
\vspace{-12pt}
\end{table*}

\begin{figure*}[t!]
  \centering
  \includegraphics[width=\linewidth]{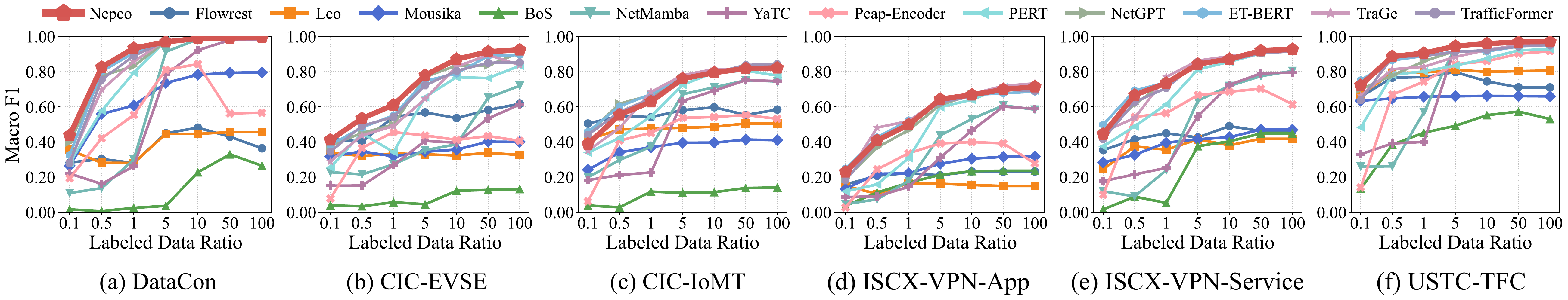}
  \caption{The classification performance comparison under limited labeled data.}
  \vspace{-14pt}
  \label{fig:Few-Shot}
\end{figure*}

\noindent\textbf{Metrics.} We primarily utilize precision, recall, and macro F1 as evaluation metrics, as they are well-suited for reflecting model performance across multi-class analysis tasks and are widely adopted in existing studies~\cite{TrafficFormer-SP25,MM4Flow-CCS25}.
All metrics are computed using Macro Averaging \cite{liu2017efficient} to prevent biased evaluation caused by class imbalance across traffic categories.

\subsection{Versatility Evaluation}
\label{sec:Versatility Evaluation}

\noindent\textbf{Overall Classification Performance.}
Compared to existing network foundation models, \textit{Nepco} achieves consistently competitive performance across all six traffic analysis tasks, as shown in TABLE \ref{tab:Overall Classification Performance}, where bold red denotes the best performance among existing network foundation models and bold blue denotes the best among existing hardware-offloadable models.
On DataCon, CIC-EVSE, CIC-IoMT, and USTC-TFC, \textit{Nepco} outperforms existing network foundation models, improving macro F1 by 1.11\% to 29.43\%.
On ISCX-VPN-App and ISCX-VPN-Service, \textit{Nepco} shows slightly lower performance (within 2\%), as several existing models, such as TrafficFormer~\cite{TrafficFormer-SP25} and NetGPT~\cite{NetGPT-arXiv23}, include these datasets in their pre-training corpora, giving them a data exposure advantage.
\cg{Under the constrained early-stage analysis setting, \textit{Nepco} delivers substantial performance improvements over existing hardware-offloadable models on the six analysis tasks.}
On average, \textit{Nepco} improves macro F1 by 160.33\% over the existing hardware-offloadable models.
Specifically, \textit{Nepco} improves over Flowrest by 116.03\%, over Leo by 149.30\%, over Mousika by 99.54\%, and over BoS by 276.46\%.
\cg{We further show in Appendix~\ref{app:Supporting Materials} that \textit{Nepco} achieves competitive classification performance with existing network foundation models on the three more network traffic datasets.}


\noindent\textbf{Labeled Data Dependency.}
We evaluate the labeled data dependency of \textit{Nepco} and baselines, by varying the labeled data ratio across 0.1\%, 0.5\%, 1\%, 5\%, 10\%, 50\%, and 100\%.
\cg{Consistent with Section \ref{sec:Motivation}, we use 5,000 flows per class and the early-stage analysis setting based on the first 5 packets of each flow.}
Concretely, a ratio of 1\% corresponds to 50 labeled flows per class, which constitutes a genuinely few-shot setting.

As shown in Fig.~\ref{fig:Few-Shot}, \textit{Nepco} achieves competitive few-shot performance compared to existing network foundation models, while existing hardware-offloadable models degrade substantially under label-scarce conditions.
Taking DataCon as an example, \textit{Nepco} achieves a macro F1 of 0.8260 under the 1\% labeled data setting, outperforming TrafficFormer (0.7549) and ET-BERT (0.8016). 
Among hardware-offloadable baselines, the best-performing Mousika reaches only 0.5593 under the 1\% labeled data setting, and even Mousika trained on the 100\% labeled data (0.7965) falls short of \textit{Nepco} trained on only 1\% of the data (0.8260).

\begin{figure}[t!]
  \centering
  \vspace{-2pt}
  \includegraphics[width=\linewidth]{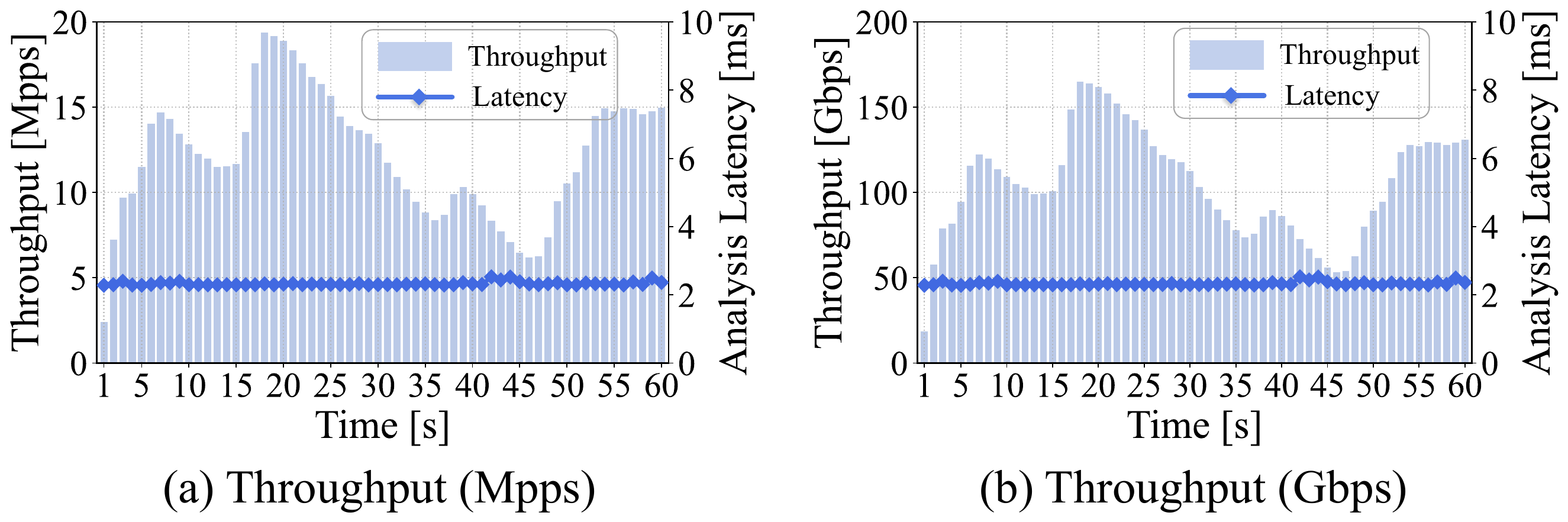}
  \caption{\cg{The analysis latency of \textit{Nepco} under continuous evaluation with varying network throughput on the SmartNIC.}}
  \vspace{-8pt}
  \label{fig:Throughput}
\end{figure}

\begin{figure}[t!]
  \centering
  \includegraphics[width=\linewidth]{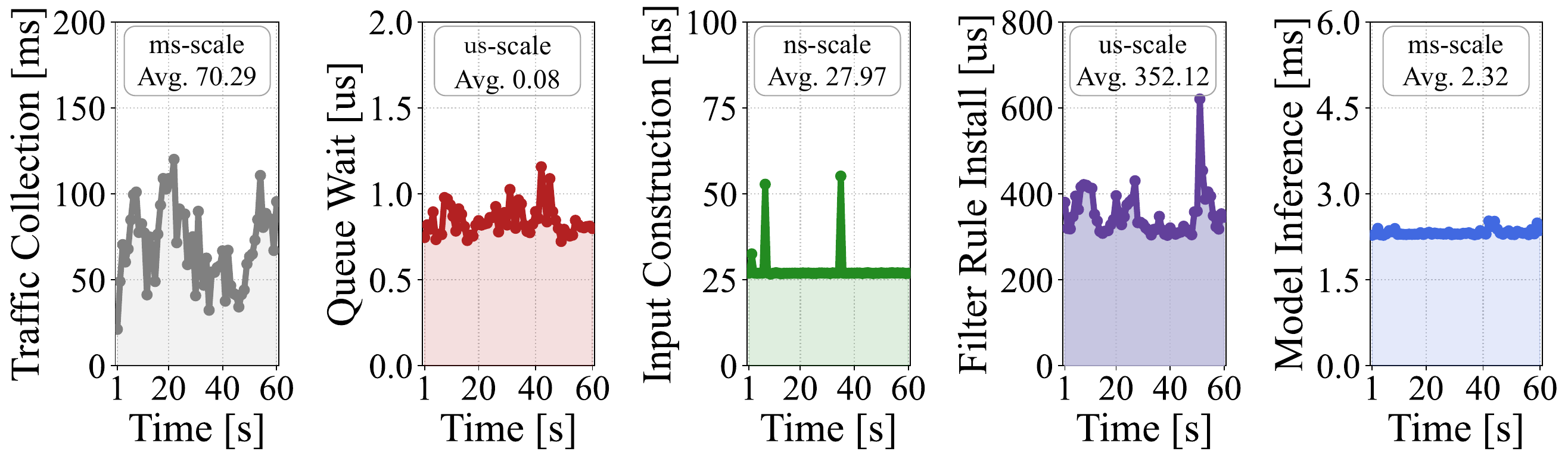}
  \caption{\cg{Latency Breakdown of \textit{Nepco} on the SmartNIC, showing traffic collection, queue wait, input construction, filter rule installation, and model inference from left to right.}}
  \vspace{-8pt}
  \label{fig:Breakdown}
\end{figure}

\begin{figure}[t!]
  \centering
  \includegraphics[width=\linewidth]{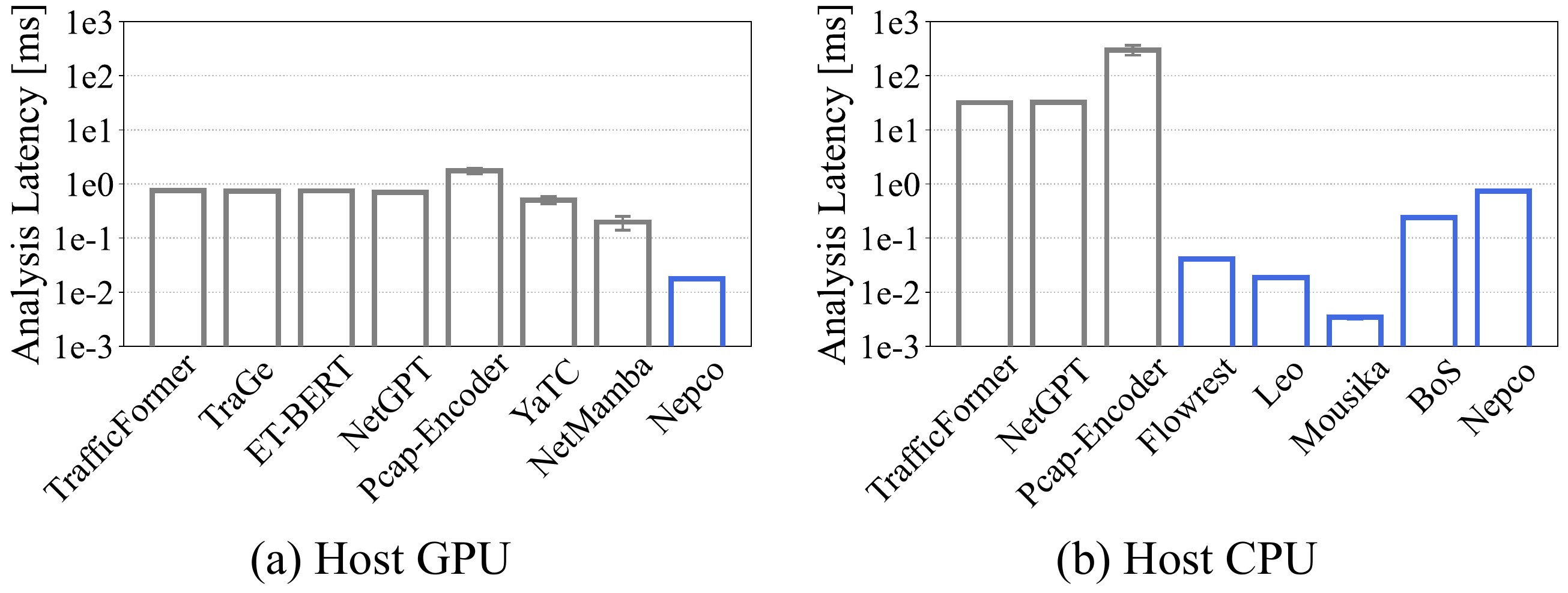}
  \caption{\cg{Analysis latency comparison under host GPU and CPU.}}
  \vspace{-10pt}
  \label{fig:Host}
\end{figure}

\subsection{Efficiency Evaluation}
\label{sec:Efficiency Evaluation}

\noindent\textbf{Throughput and Latency.}
We evaluate the hardware prototype of \textit{Nepco} deployed on the SmartNIC under continuous traffic over a one-minute measurement, reporting per-second average throughput and analysis latency. 
\cg{The analysis latency represents the total time from the end of traffic collection to the end of model inference, following existing work \cite{BoS-NSDI24,NetBeacon-USENIX-Security23}}.
As shown in Fig.~\ref{fig:Throughput}, under network 
throughput reaching up to 19.81~Mpps and 165.14~Gbps, 
\textit{Nepco} maintains a stable analysis latency between 2.23~ms and 2.47~ms, demonstrating efficient traffic analysis under high-speed network conditions.

We further analyze the latency breakdown of \textit{Nepco} in Fig.~\ref{fig:Breakdown}.
\cg{Traffic collection and model inference operate at the ms-scale, filter rule installation and queue wait at the $\mu$s-scale, and input construction at the ns-scale, with average latencies of 70.29~ms, 2.32~ms, 352.12~$\mu$s, 0.08~$\mu$s, and 27.97~ns, respectively.
Based on the breakdown, the first-packet-to-decision latency, first-packet-to-filter-rule-installed latency, and fifth-packet-to-decision latency average 72.61~ms, 70.64~ms, and 2.32~ms, respectively.
Specifically, the first-packet-to-decision latency includes traffic collection, queue wait, input construction, and model inference, whereas the first-packet-to-filter-rule-installed latency includes traffic collection and filter-rule installation.
The fifth-packet-to-decision latency corresponds to the analysis latency reported above.}

\cg{We evaluate the efficiency of \textit{Nepco}'s architecture design. For this, we evaluate the software prototypes of \textit{Nepco} and the baselines on host GPU and CPU platforms, as shown in Fig.~\ref{fig:Host}.}
On the host GPU, \textit{Nepco} strictly outperforms existing network foundation models, maintaining an average analysis latency of 0.0178~ms and reducing analysis latency by 11.10$\times$ to 98.46$\times$.
On the host CPU, \textit{Nepco} further demonstrates its lightweight nature with an average analysis latency of 0.7395~ms, reducing analysis latency by 43.27$\times$ to 408.37$\times$. 
\cg{Overall, \textit{Nepco}'s architectural design reduces analysis latency by up to 98.46$\times$ on the host GPU, independent of the SmartNIC deployment.}

\begin{table}[t]
\renewcommand{\arraystretch}{1}
\centering
\small
\caption{\cg{Comparison results of \textit{Nepco} and existing foundation models on SmartNICs under different network loads.}}
\label{tab:bf3-nepco-results}
\resizebox{0.98 \columnwidth}{!}{%
\begin{tabular}{c|c|c|c|c}

\toprule
\makecell{\textbf{Network Load}} &
\makecell{\textbf{Method}} &
\makecell{\textbf{Finish} \\ \textbf{Flow Num.}} &
\makecell{\textbf{Avg. Analysis} \\ \textbf{Latency [ms]}} &
\makecell{\textbf{P99 Analysis} \\ \textbf{Latency [ms]}} \\
\midrule

\multirow{4}{*}{\makecell{\textbf{Extre. Low} \\ NFPS = 5}} &
ET-BERT &
300 &
786.57 &
881.37 \\

 &
TraGe &
300 &
781.36 &
825.17 \\

 & 
TrafficFormer &
300 & 
791.50 &
887.91 \\ \cmidrule(lr){2-5}

 & 
\textit{\textbf{Nepco}} & 
\textbf{300} & 
\textbf{2.31} &
\textbf{2.61} \\

\midrule
\multirow{4}{*}{\makecell{ \textbf{Low} \\ NFPS = 500}} &
ET-BERT &
981 &
30479.24 &
60242.34 \\

 & 
TraGe &
966 &
30694.34 & 
60493.45 \\

 &
TrafficFormer & 
985 &
30548.47 & 
60106.11 \\ \cmidrule(lr){2-5}

 &
\textit{\textbf{Nepco}} & 
\textbf{30,000} &
\textbf{2.36} &
\textbf{2.68} \\
\midrule

\multirow{4}{*}{\makecell{\textbf{Normal} \\ NFPS = 2000}} & 
ET-BERT &
982 &
31532.24 &
62117.65 \\

 & 
TraGe &
1,007 &
31890.26 &
62889.06 \\

 & 
TrafficFormer &
998 &
31957.56 &
63105.05 \\ \cmidrule(lr){2-5}

 & 
\textit{\textbf{Nepco}} & 
\textbf{120,000} &
\textbf{2.38} & 
\textbf{2.69} \\

\midrule
\multirow{4}{*}{\makecell{\textbf{High} \\ NFPS = 5000}} &
ET-BERT &
1,053 &
33619.74 &
65698.44 \\

&
TraGe &
992 &
31944.91 &
62351.57 \\

& 
TrafficFormer &
1,039 &
33027.27 &
64730.15 \\ \cmidrule(lr){2-5}

& 
\textit{\textbf{Nepco}} & 
\textbf{300,000} & 
\textbf{2.41} &
\textbf{2.71} \\

\bottomrule

\end{tabular}%
}
\vspace{-4pt}
\end{table}

\begin{table}[t]
\renewcommand{\arraystretch}{1}
\centering
\small
\caption{\cg{Resource utilization comparison on the SmartNIC.}}
\label{tab:resource-utilization}
\resizebox{0.98\columnwidth}{!}{%
\begin{tabular}{c|c|c|c|c|c|c|c}

\toprule

\multirow{2}{*}{\textbf{Method}} &
\multirow{2}{*}{
    \begin{tabular}[c]{@{}c@{}}
    \textbf{Power}\\
    \textbf{[W]}
    \end{tabular}
} &
\multirow{2}{*}{
    \begin{tabular}[c]{@{}c@{}}
    \textbf{ARM}\\
    \textbf{Core}
    \end{tabular}
} &
\multirow{2}{*}{
    \begin{tabular}[c]{@{}c@{}}
    \textbf{Lookup}\\
    \textbf{Table}
    \end{tabular}
} &
\multicolumn{4}{c}{\textbf{Memory [MB] \textsuperscript{1}}} \\

& & & &
\textbf{R. O.} &
\textbf{M. F.} &
\textbf{I. M.} &
\textbf{Total} \\

\midrule

ET-BERT &
68 &
15 &
50k &
58.69 &
728.12 &
67.88 &
854.69 \\

TraGe &
68 &
15 &
50k &
58.75 &
728.06 &
60.88 &
847.69 \\

TrafficFormer &
69 &
15 &
50k &
58.69 &
728.12 &
60.25 &
847.06 \\

\cmidrule(lr){1-8}

\textit{\textbf{Nepco}} &
\textbf{64} &
\textbf{15} &
\textbf{50k} &
\textbf{58.90} &
\textbf{40.50} &
\textbf{17.10} &
\textbf{116.50} \\

\bottomrule

\multicolumn{8}{l}{\textsuperscript{1} \textbf{R}untime \textbf{O}verhead, \textbf{M}odel \textbf{F}ootprint, and \textbf{I}nference \textbf{W}orkspace} \\

\end{tabular}%
}
\vspace{-12pt}
\end{table}

\noindent\textbf{Diverse Network Loads.}
\cg{We evaluate the efficiency of \textit{Nepco}'s SmartNIC deployment under diverse network loads. For this, we deploy \textit{Nepco} and three existing network foundation models on the same SmartNIC and measure their analysis latency under network loads characterized by new flows per second (NFPS).}
In TABLE~\ref{tab:bf3-nepco-results}, existing network foundation models achieve line-rate analysis under low network load (e.g., TrafficFormer reaches 791.50 ms at 5 NFPS), but their latency increases significantly as the network load grows.
In contrast, \textit{Nepco} maintains millisecond-scale analysis latency across four network loads, and achieves line-rate analysis even under high load (5,000 NFPS).
\cg{Notably, at 5 NFPS, where both \textit{Nepco} and existing models can process the arriving flows at line rate, \textit{Nepco} reduces analysis latency by up to 328$\times$.}
\cg{Overall, under the same SmartNIC deployment setting, \textit{Nepco} reduces the analysis latency of existing network foundation models from the second scale to the millisecond scale, achieving up to 328$\times$ analysis latency reduction.
Together with Fig. \ref{fig:Host}, these results distinguish the architectural efficiency of \textit{Nepco} from its system-level efficiency under SmartNIC deployment.}

\noindent\textbf{\cg{Hardware Resource Utilization.}}
\cg{We further report the hardware resource utilization on the SmartNIC in TABLE~\ref{tab:resource-utilization}. Under the same hardware configuration of 15 SoC ARM cores and 50k NIC lookup-table entries, \textit{Nepco} consumes less power and memory than the existing network foundation models.
The BlueField-3 SmartNIC has a maximum power budget of 150~W and consumes 32~W when idle. \textit{Nepco} reaches a peak power consumption of 64~W during execution, compared with 68~W for ET-BERT and TraGe and 69~W for TrafficFormer. We wait for the SmartNIC temperature to return to normal before each power measurement to avoid temperature-induced power variation.
For memory, \textit{Nepco} consumes 58.90~MB for runtime overhead (i.e., software runtime and supporting data structures), 40.50~MB for model footprint (i.e., model and inference session), and 17.10~MB for inference workspace (i.e., intermediate tensors and runtime workspace), totaling 116.50~MB. In comparison, ET-BERT, TraGe, and TrafficFormer consume 854.69~MB, 847.69~MB, and 847.06~MB, respectively. Overall, \textit{Nepco} reduces memory usage by 7.27$\times$ to 7.34$\times$ under the SmartNIC deployment setting.}

\subsection{Nepco Deep Dive}
\label{sec:Nepco Deep Dive}

\begin{figure}[t]
  \centering
  \includegraphics[width=\linewidth]{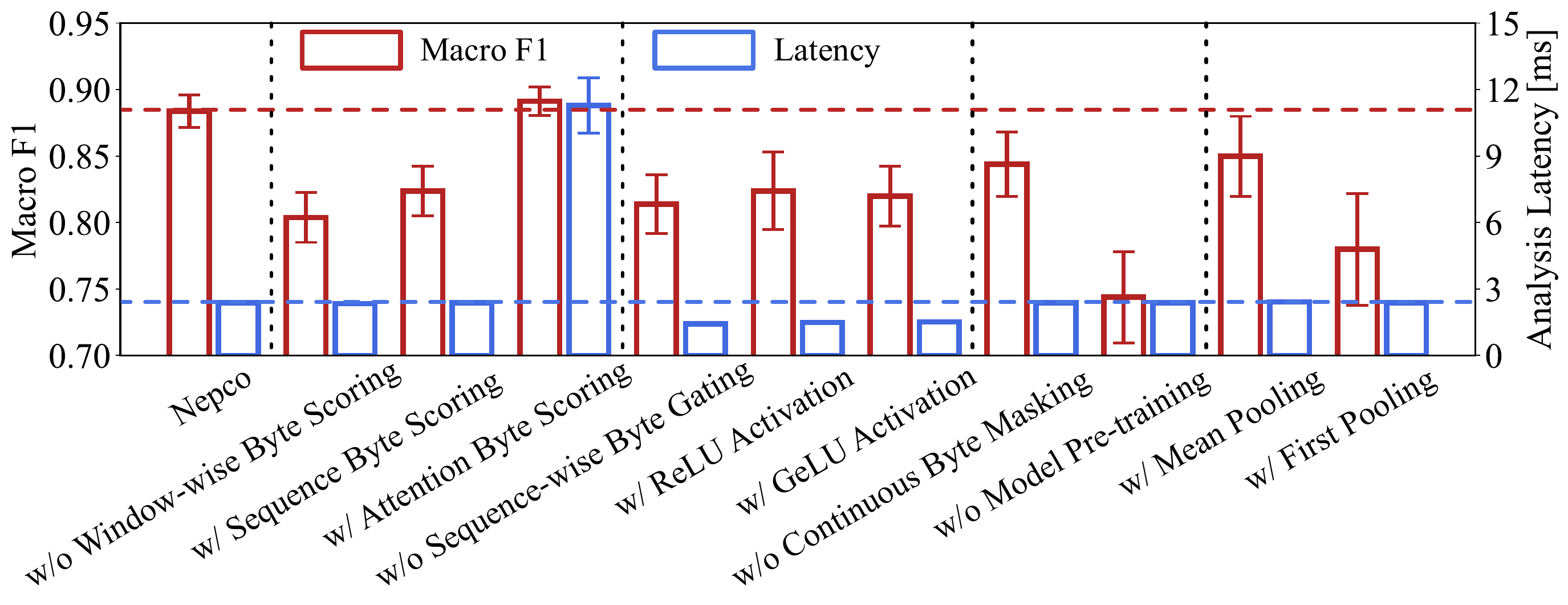}
  \caption{\cg{Ablation study of \textit{Nepco}, including the key component ablation (w/o component) and the substitution with existing methods (w/ component replaced by alternative).}}
  \vspace{-6pt}
  \label{fig:Ablation}
\end{figure}

\begin{figure}[t]
  \centering
  \includegraphics[width=\linewidth]{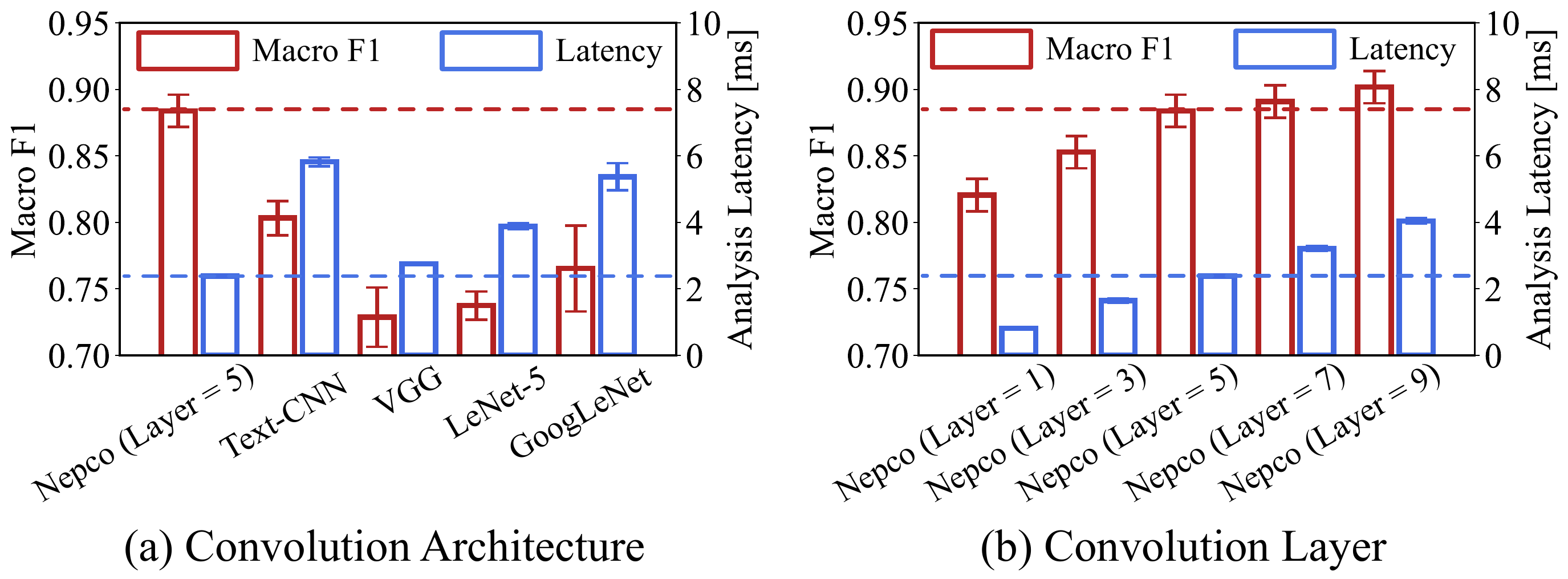}
  \caption{\cg{The comparison with convolution-based methods.}}
  \vspace{-4pt}
  \label{fig:Conv_Arch}
\end{figure}

\begin{figure}[t]
  \centering
  \includegraphics[width=\linewidth]{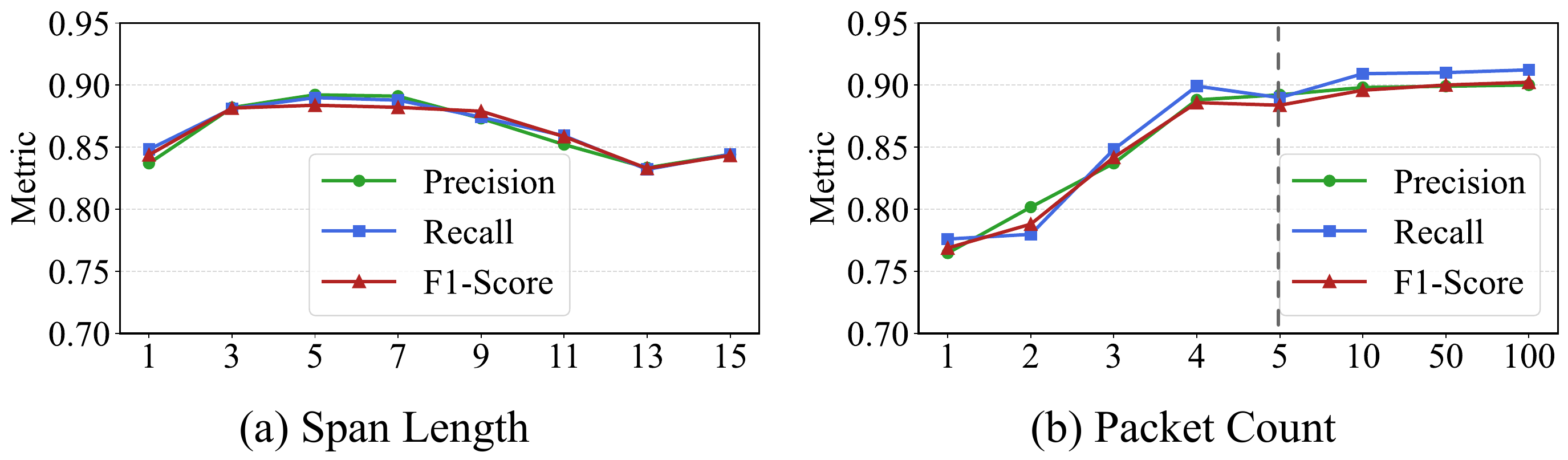}
  \caption{Parameter sensitivity analysis of \textit{Nepco}.}
  \vspace{-12pt}
  \label{fig:Para_Sen}
\end{figure}

\noindent\textbf{Ablation Study.}
We validate the model architecture design of \textit{Nepco} in Fig.~\ref{fig:Ablation}.
To begin with, removing window-wise byte scoring leads to a 9.05\% drop in macro F1, while replacing it with sequence-level scoring yields only marginal improvement, \cg{indicating that explicitly modeling localized patterns within each convolution window provides substantial benefits beyond sequence-level scoring.}
\cg{Although attention-based scoring slightly improves performance (+0.85\%), it incurs a 4.74$\times$ increase in analysis latency, showing that the proposed design achieves a better efficiency–effectiveness trade-off.}
Consistent with this, removing sequence-wise byte gating causes an 8.71\% drop, and replacing it with ReLU or GeLU brings limited gains, suggesting that adaptive importance estimation is necessary beyond static activations. 
From the training perspective, replacing continuous byte masking with random masking significantly degrades performance, and removing pre-training further leads to a 15.84\% drop, showing the importance of learning generalizable traffic representations. 
Finally, replacing max pooling with mean or first pooling degrades performance by 3.85\% and 11.77\%, respectively, as they fail to preserve discriminative localized patterns.

\noindent\textbf{Convolution-based Methods.}
We compare \textit{Nepco} against representative convolution-based methods, including Text-CNN \cite{Text-CNN-EMNLP14}, VGG \cite{VGG-arXiv14}, LeNet-5 \cite{LeNeet-5-IEEE02}, and GoogLeNet \cite{GoogLeNet-CVPR15}, which are widely used as model architectures in existing traffic analysis methods \cite{Wang-ISI17, USTC-TFC-ICOIN17, CETAnalytics-CN20, ICLSTM-MDPI21, HexCNN-JCC23, Yu-CNSCT25}. 
As shown in Fig.~\ref{fig:Conv_Arch}(a), \textit{Nepco} outperforms existing convolution-based methods by 10.03\%, 21.28\%, 19.85\%, and 15.48\% in macro F1, demonstrating the effectiveness of \textit{Nepco}'s localized pattern-aware design.
We further analyze the effect of the number of traffic convolution layers in Fig.~\ref{fig:Conv_Arch}(b). The results show that 5 layers achieve the best trade-off between macro F1 and analysis latency.
Fewer layers reduce representation capacity, while more layers increase analysis latency linearly without substantial performance gains.

\noindent\textbf{Parameter Sensitivity Analysis.}
\cg{We analyze the sensitivity of two key parameters in \textit{Nepco} on the CIC-EVSE dataset.}
Fig.~\ref{fig:Para_Sen}(a) shows the effect of the span length in continuous byte masking.
A span length between 3 and 7 achieves the best performance, as shorter spans degrade to random masking and fail to capture localized protocol field patterns, while longer spans increase learning difficulty and deviate from the byte length distribution of protocol fields.
Fig.~\ref{fig:Para_Sen}(b) shows the effect of packet count on classification 
performance. Increasing the packet count below 5 yields substantial gains, while further increasing from 5 to 10, 50, or even 100 provides diminishing returns. This confirms that fine-grained byte-level semantics reduce reliance on observing more packets, consistent with the early-stage analysis setting adopted by prior works~\cite{TrafficFormer-SP25}.

%% file: Discussion.tex
\section{Discussion}
\label{sec:Discussion}

\noindent\textbf{Pre-training Corpus Scale and Generalizability.}
The pre-training corpus of \textit{Nepco} is moderate in scale, in part because we strictly require the pre-training data to predate the fine-tuning data.
This temporal separation avoids using future protocol knowledge to evaluate earlier analysis tasks, which would inflate generalization evaluation.
Despite this constraint, Section \ref{sec:Versatility Evaluation} shows that \textit{Nepco} achieves classification performance competitive with state-of-the-art network foundation models.
Recent work, such as MM4Flow \cite{MM4Flow-CCS25}, shows that TB-scale pre-training further enhances versatility, which we leave as an orthogonal direction for future work.

\noindent\textbf{\cg{Adversarial Robustness.}}
\cg{\textit{Nepco} builds on the observation that discriminative traffic information often exhibits localized patterns within byte sequences, and accordingly performs efficient localized pattern-aware traffic modeling.
Such localized pattern-aware traffic modeling may render model decisions more sensitive to targeted perturbations of a limited set of discriminative bytes than Transformer-based network foundation models that perform global sequence modeling.
We therefore evaluate the adversarial robustness of \textit{Nepco} against byte-level perturbations on CIC-EVSE, in comparison with Transformer-based network foundation models (i.e., ET-BERT, TraGe, and TrafficFormer).
Specifically, we construct two byte-level perturbation strategies. Random-X\% perturbation randomly selects X\% of the input bytes, whereas Top-X\% perturbation selects the top X\% most important bytes according to gradient-based attribution \cite{gradient-arXiv17}.
The selected bytes are randomized under both perturbation strategies.
We train each model on the raw training set and evaluate it on test sets perturbed by Random-X\% and Top-X\% perturbations at different values of X.}

\cg{
The results show that \textit{Nepco} exhibits adversarial robustness comparable to the Transformer-based network foundation models under both perturbation strategies.
Specifically, at a 5\% perturbation ratio, the relative macro F1 reductions of \textit{Nepco}, TrafficFormer, TraGe, and ET-BERT are 5.57\%, 8.02\%, 5.49\%, and 2.11\%, respectively, under Random-5\% perturbation, and 49.67\%, 53.16\%, 54.19\%, and 54.00\%, respectively, under Top-5\% perturbation.
These results suggest that \textit{Nepco}'s localized pattern-aware modeling does not exhibit greater vulnerability to targeted important-byte perturbations than Transformer-based global modeling.
To further mitigate the impact of such perturbations, we retrain each model on training sets perturbed by Random-X\% and Top-X\% perturbations at the corresponding values of X, and evaluate the retrained models on the corresponding perturbed test sets.
At X=5\%, under Top-5\% perturbation, the macro F1 of \textit{Nepco}, TrafficFormer, TraGe, and ET-BERT increases from 0.4451, 0.4190, 0.4018, and 0.4057 to 0.8447, 0.9082, 0.8987, and 0.8918, respectively.
These results show that training with correspondingly perturbed traffic samples can effectively mitigate the impact of byte-level perturbations.
Detailed experimental settings and results are provided in Appendix~\ref{app:Supporting Materials}.}

\noindent\textbf{Deployment on Other Programmable Hardware.}
\textit{Nepco} is prototyped on the SmartNIC, but its design choices can be extended to other programmable hardware platforms.
Specifically, \textit{Nepco}'s window-wise byte scoring uses pre-computed weights that can be realized as table lookups, while its sequence-wise byte gating relies on parallel element-wise operations—both are inherently hardware-friendly. This allows \textit{Nepco} to align with parallel-compute platforms such as FPGAs and to enable hybrid execution on more constrained platforms such as P4 switches.
We believe \textit{Nepco}'s core design (e.g., efficient localized byte-sequence modeling) has the potential to inform and strengthen network foundation model-based traffic analysis on other programmable hardware platforms.

\noindent\textbf{\cg{SmartNIC Deployment of LLM-based Foundation Models.}}
\cg{We further examine the feasibility of applying \textit{Nepco}'s SmartNIC deployment scheme to a recent Large Language Model (LLM) based network foundation model, i.e., TrafficLLM \cite{TrafficLLM-arXiv25}.
We attempt to deploy TrafficLLM under the same SmartNIC deployment scheme of \textit{Nepco}, but TrafficLLM's complete model inference runtime cannot fit within the 32-GB memory capacity of the BlueField-3 SmartNIC.
This is mainly because TrafficLLM adopts open-source large language models, i.e., Llama2-7B \cite{LLAMA2-arxiv23} and ChatGLM2-6B \cite{ChatGLM-arxiv22}, as its base LLMs, whose billion-parameter scale imposes substantial memory demands.
Moreover, according to the original TrafficLLM study, deploying TrafficLLM on a host GPU (i.e., NVIDIA A100-80GB) requires approximately 13 GB of GPU memory for model loading and incurs around 0.2 s of model inference latency per flow.
Overall, fully offloading LLM-based network foundation models such as TrafficLLM to SmartNICs requires further reducing their inference memory footprint, e.g., through model compression or smaller base LLMs.}

%% file: Conclusion.tex
\section{Conclusion}
\label{sec:Conclusion}

\cg{In this paper, we show that the long-standing challenge between versatile analysis with network foundation models and efficient analysis with hardware offloading is not fundamental, but a consequence of polarized design choices in prior work.}
We present \textit{Nepco}, which resolves these polarizations and offloads the network foundation model to SmartNIC for versatile yet efficient network traffic analysis.
\cg{At its core, we co-design the traffic analysis system through hardware-friendly traffic processing, observation-driven convolution architecture, and SmartNIC-native collaborative execution.}
Across six traffic analysis tasks, \textit{Nepco} matches or exceeds state-of-the-art network foundation models, \cg{while reducing analysis latency by 328$\times$ to the millisecond scale under the SmartNIC deployment setting.}
\cg{Under the early-stage analysis setting, \textit{Nepco} trained with only 1\% of labeled data surpasses existing hardware-offloadable models trained with 100\% of labeled data.}

%% file: Acknowledgment.tex
\section*{Acknowledgment}

We thank the anonymous reviewers for their constructive feedback.
This work was supported in whole or in part by the National Natural Science Foundation of China (U24B6012, 62372429, U2333201, U23A20304, and U25A20428), the Pilot for Major Scientific Research Facility of Jiangsu Province of China (No.BM2021800), and the Beijing Natural Science Foundation Project with No. Z260008.

%% file: Appendix.tex

\input{Appendix/Support-Short}
\clearpage

\input{Appendix/Artifact}

%% file: Appendix/Support-Short.tex
\section{Supporting Materials}
\label{app:Supporting Materials}

Due to space limitations, we provide supporting materials on Zenodo \cite{Nepco-Appendix}. For a complete presentation of the study, these materials can be downloaded and combined with the paper.


\subsection{The Impact of Address-related Information}
We analyze the impact of address-related information on network foundation models during pre-training and fine-tuning. We compare models with and without address-related pre-training tasks and evaluate the effects of retaining, zeroing out, and randomizing address fields during fine-tuning evaluation. The results show that address information inflates classification performance, while test-time randomization causes substantial performance degradation, supporting the use of address-field zeroing to reduce address-related shortcuts.

\subsection{Cross-Dataset Localized Byte-sequence Pattern}
We analyze byte and field importance on USTC-TFC and CIC-IoT to provide cross-dataset support for the observations motivating our design. We quantify byte importance concentration, class-specific field preferences, within-field importance concentration, and address-field importance. The results show that importance concentrates on subsets of bytes, different classes favor different field regions, and address fields receive disproportionate importance, supporting localized byte-sequence modeling and address-field zeroing.

\subsection{Evaluation on Three More Network Traffic Datasets}
We evaluate \textit{Nepco} and baselines on three more network traffic datasets (i.e., CSTNET-20, CSTNET-120, and DoHBrw~\cite{DoHBrw-DASC20}), covering TLS 1.3-encrypted mobile application traffic and malicious DNS-over-HTTPS traffic. \textit{Nepco} achieves the highest macro F1 among the evaluated network foundation models on CSTNET-20 and CSTNET-120, while performing slightly below the best performance achieved by existing network foundation models on DoHBrw. These results demonstrate that \textit{Nepco} achieves competitive classification performance with state-of-the-art network foundation models under limited labeled data.

\subsection{Adversarial Robustness Evaluation}
We evaluate the adversarial robustness of \textit{Nepco} on the CIC-EVSE dataset against three Transformer-based network foundation models, i.e., ET-BERT, TraGe, and TrafficFormer. Under random and importance-guided byte perturbations, the four models are more sensitive to perturbations targeting high-importance bytes, with \textit{Nepco} exhibiting comparable sensitivity. We further explore training with perturbed samples as a mitigation strategy. Incorporating corresponding perturbations into training substantially reduces performance degradation, with the four models maintaining macro F1 above 0.84 across the evaluated perturbation ratios.

%% file: Appendix/Artifact.tex
\section{Artifact Appendix}
\label{app:Artifact}

This artifact provides the implementation and evaluation code for \textit{Nepco}, the hardware-offloadable network foundation model.
It includes complete workflows for data processing, model training, host and SmartNIC deployment, and adversarial evaluation, together with environment setup scripts, test datasets, model checkpoints, and detailed instructions for reproducing the main results presented in the paper.

\subsection{Description \& Requirements}

\subsubsection{How to access}
The artifact is available on GitHub at \url{https://github.com/lincgcg/Nepco}, with an archived version available on Zenodo at \url{https://zenodo.org/records/22893365} under DOI \url{https://doi.org/10.5281/zenodo.22893365}.
The implementation, configuration, vocabulary, and pre-trained model are under \texttt{uer/}, \texttt{configs/}, \texttt{vocab/}, and \texttt{models/}. The two evaluation workflows are under \texttt{E1/} and \texttt{E2/}. Each experiment includes a README and manifest. 

\subsubsection{Hardware dependencies}
For model pre-training and fine-tuning, we recommend a Linux machine with at least 32~GB RAM and a CUDA-capable NVIDIA GPU. Our experiments used an NVIDIA H800 GPU with 80~GB VRAM and an Intel Xeon Platinum 8458P CPU.
For host-side deployment and latency measurement, CPU execution is supported, while GPU evaluation requires a CUDA-capable NVIDIA GPU.
SmartNIC evaluation requires two NVIDIA BlueField-3 SmartNICs; our testbed uses Intel Core i7-14700 hosts with 64~GB RAM and a 200~Gbps link.

To enable lightweight evaluation without retraining or specialized SmartNIC hardware, we provide E1 and E2 with CPU-based execution paths. These require eight available logical CPU cores, for which we recommend 8~GB RAM and sufficient disk space for the downloaded datasets, checkpoints, and generated results. GPU evaluation in E2 additionally requires a CUDA-capable NVIDIA GPU.

\subsubsection{Software dependencies}
Python dependencies are listed in \texttt{requirements.txt}. The verified environment uses Python~3.10.14 and PyTorch~2.3.0, with CUDA~12.4 for GPU execution. CPU latency measurement additionally requires Linux \texttt{taskset}. SmartNIC deployment requires NVIDIA DOCA/DPDK, GNU Make, a C++ compiler, and ONNX Runtime for C/C++. No Internet access is required during execution once dependencies and artifacts are available locally.


\subsubsection{Benchmarks}
We provide labeled test datasets and fine-tuned checkpoints for nine traffic-analysis datasets: DataCon, CIC-EVSE, CIC-IoMT, ISCX-VPN-App, ISCX-VPN-Service, USTC-TFC, CSTNET-20, CSTNET-120, and DoHBrw. The artifacts include test TSV files, \textit{Nepco} and TraGe checkpoints, and their model configurations, supporting classification evaluation and host-side inference-latency measurement.
Since their total size exceeds 5~GB, these files are distributed through Google Drive at \url{https://drive.google.com/drive/folders/1UafVUJeugAGDHBASkYFizduFhkPXuiBF}. The artifact contents, directory structure, and instructions for downloading and placing the files are provided in \path{E1/artifacts/README.md}.



\subsection{Artifact Installation \& Configuration}

\noindent\textit{[Environment Setup]}
From the repository root, create a Python environment and install the dependencies:
\begin{verbatim}
python3 -m venv .venv
source .venv/bin/activate
python3 -m pip install -r requirements.txt
\end{verbatim}

\noindent\textit{[Artifact Download]}
Download the benchmark artifacts from the Google Drive link above and place the downloaded dataset directories under \path{E1/artifacts/}, preserving their original directory structure. Confirm that the required test TSV files, model checkpoints, and model configurations are present, together with \path{configs/nepco_config.json} and \path{vocab/hex_vocab.txt}.

\subsection{Experiment Workflow and Claims}

The artifact provides five functional workflows covering input data processing, model training, host-side latency measurement, SmartNIC-side latency measurement, and adversarial robustness evaluation.
We organize the evaluation into two experiments addressing the core aspects of \textit{Nepco}: versatility (E1), evaluated through classification performance across nine traffic-analysis datasets, and efficiency (E2), evaluated through traffic-analysis latency under host and SmartNIC deployment.

\begin{itemize}
    \item \textbf{(C1) Versatility.}
    Compared with existing network foundation models, \textit{Nepco} achieves competitive classification performance.
    This claim is supported by experiment (E1).
    Specifically, E1.1 evaluates \textit{Nepco} on nine traffic-analysis datasets, while E1.2 evaluates an existing network foundation model, TraGe, on the same datasets.
    These datasets correspond to those reported in TABLE~III and TABLE~VI.

    \item \textbf{(C2) Efficiency.}
    Compared with existing network foundation models, \textit{Nepco} achieves lower traffic-analysis latency.
    This claim is supported by experiment (E2).
    Specifically, E2.1 measures the traffic-analysis latency of \textit{Nepco} and TraGe under host CPU and GPU deployment, corresponding to Fig.~9.
    E2.2 measures \textit{Nepco}'s traffic-analysis latency under SmartNIC deployment, corresponding to Fig.~7, Fig.~8, and TABLE~IV.
    E2.2 requires two NVIDIA BlueField-3 SmartNICs and the associated host and network infrastructure, while E2.1 supports C2 through host-side deployment without requiring SmartNIC hardware.
\end{itemize}

\subsection{Evaluation}

\subsubsection{Experiment (E1): Versatility Evaluation}

This experiment evaluates the classification performance of \textit{Nepco} and TraGe across the nine traffic-analysis datasets reported in TABLE~III and TABLE~VI, supporting C1. The supplied fine-tuned checkpoints are evaluated without retraining.

\noindent\textbf{\textit{E1.1: Nepco Classification Performance}}

Required time: less than one human-minute after setup; compute time varies by hardware.

\noindent\textit{[Preparation]}
Download and arrange the benchmark artifacts as described in the installation section. Each dataset directory under \path{E1/artifacts/} should contain \texttt{test\_dataset.tsv}, \texttt{finetuned\_model.bin}, and \texttt{model\_config.json}.

\noindent\textit{[Execution]}
From the repository root, run:
\begin{verbatim}
bash E1/run_smoke_test.sh --all
\end{verbatim}
The script checks artifact hashes, loads the corresponding checkpoint and configuration for each dataset, and evaluates the complete test set. CPU execution is the default; append \texttt{--device cuda} for GPU execution.

\noindent\textit{[Results]}
The evaluator reports macro precision, recall, and F1, and saves \texttt{metrics.json}, \texttt{prf.csv}, and \texttt{predictions.tsv} under \path{E1/results/<dataset>/reproduced/}. Confirm that each complete test set is processed and the prediction count matches the sample count. Compare the metrics with the archived results under \path{E1/results/<dataset>/} and the \textit{Nepco} entries in TABLE~III and TABLE~VI. For example, the supplied DataCon and USTC-TFC checkpoints yield macro F1 scores of 0.991604 and 0.963832, respectively.

\noindent\textbf{\textit{E1.2: TraGe Classification Performance}}

Required time: less than one human-minute after setup; compute time varies by hardware.

\noindent\textit{[Preparation]}
Use the test TSV files from E1.1. Confirm that each dataset directory contains the TraGe checkpoint and configuration under \path{E1/artifacts/<dataset>/TraGe/}.

\noindent\textit{[Execution]}
From the repository root, run:
\begin{verbatim}
bash E1/run_trage.sh all
\end{verbatim}
The script checks artifact hashes and evaluates the supplied TraGe checkpoints on nine complete test sets. CPU execution is default; append \texttt{--device cuda} for GPU execution.

\noindent\textit{[Results]}
The evaluator reports the same metrics as E1.1, saves \texttt{metrics.json}, \texttt{prf.csv}, and \texttt{predictions.tsv} under \path{E1/results/<dataset>/TraGe/reproduced/}, and checks agreement with the archived confusion matrices. Compare the metrics with the archived results under \path{E1/results/<dataset>/TraGe/} and the TraGe entries in TABLE~III and TABLE~VI. For example, the supplied DataCon and USTC-TFC checkpoints yield macro F1 scores of 0.987727 and 0.958106, respectively.

\subsubsection{Experiment (E2): Efficiency Evaluation}

This experiment evaluates analysis latency under host CPU/GPU deployment (E2.1) and SmartNIC deployment (E2.2), supporting C2.

\noindent\textbf{\textit{E2.1: Host CPU/GPU Analysis Latency}}

Required time: a few human-minutes after setup; compute time varies by hardware.

\noindent\textit{[Preparation]}
Use the DataCon artifacts from E1. CPU evaluation requires Linux \texttt{taskset}; GPU evaluation requires a CUDA-capable NVIDIA GPU. The complete matrix uses up to 64 physical CPU cores.

\noindent\textit{[Execution]}
From the repository root, run:
\begin{verbatim}
bash E2/run_benchmark.sh --gpu_index 0
\end{verbatim}
The script sequentially benchmarks \textit{Nepco} and TraGe in CPU and GPU modes. Both modes use batch sizes of 1, 2, 4, 8, 16, 32, 64, and 128; CPU mode additionally varies the core count over 8, 16, 32, and 64. Each configuration processes the complete 2,857-sample test set with one warm-up pass and five measured passes. Only model-forward execution is timed, with CUDA synchronization in GPU mode.

\noindent\textit{[Results]}
Results are saved under \path{E2/results/local_runs/}, with predictions checked against E1 references. Compare per-sample latency with the archived results under \path{E2/results/DataCon/full_benchmark/} and Fig.~9. For example, at batch size~1, \textit{Nepco}/TraGe achieve 2.078/45.543~ms on eight CPU cores and 0.571/4.807~ms on GPU. Timings vary by hardware and load; batched values are amortized per sample.

\noindent\textbf{\textit{E2.2: SmartNIC Analysis Latency}}

Required time: a few human-minutes after testbed setup; one minute per measurement window.

\noindent\textit{[Preparation]}
Prepare two connected NVIDIA BlueField-3 SmartNICs with DOCA/DPDK and ONNX Runtime. Configure device access, hugepages, and CPU affinity. Export the selected checkpoint using \path{scripts/4_latency_measure_SmartNIC/1_onnx_conversion/run_onnx_export.sh}, specifying the checkpoint, calibration TSV, and class count.

\noindent\textit{[Execution]}
Run the following commands on the corresponding systems, with the model, vocabulary, ONNX Runtime, and device settings configured:
\begin{verbatim}
DEPLOY=scripts/4_latency_measure_SmartNIC/\
2_SmartNIC_deployment

# Receiver
bash "$DEPLOY/run_build_receiver.sh"
bash "$DEPLOY/run_receiver.sh"

# Sender
bash "$DEPLOY/run_build_sender.sh"
bash "$DEPLOY/run_sender.sh"
\end{verbatim}

\noindent\textit{[Results]}
The sender reports throughput; the receiver logs queue wait, input construction, inference, and total analysis latency under \texttt{OUTPUT\_DIR}. These measurements support the SmartNIC evaluation associated with Figs.~7--8 and TABLE~IV. Separate collection time when reporting post-collection analysis latency. E2.1 supports C2 without SmartNIC hardware.

\subsection{Customization and Scope}

E1 supports dataset selection and CPU/GPU execution. E2.1 supports individual model, device, batch-size, and CPU-core configurations, while E2.2 allows testbed-specific model and network settings. E1 and E2 evaluate versatility and efficiency using released checkpoints without retraining. Additional workflows for data processing, training, and adversarial evaluation are provided under \path{scripts/}; \path{E_Mapping/README.md} maps these workflows to the corresponding figures and tables in the paper.